\documentclass[a4paper,11pt]{article}
\pdfoutput=1 

\usepackage{jcappub} 
\usepackage{amsfonts}
\usepackage[utf8]{inputenc}
\usepackage{graphicx}
\usepackage{dcolumn}
\usepackage{bm}
\usepackage{amsmath}
\usepackage{lipsum}
\usepackage{xcolor}
\usepackage{calc}
\usepackage{accents}
\usepackage{float}
\usepackage{pifont}
\usepackage{makecell}
\usepackage{caption}
\usepackage{subcaption}
\usepackage{epsfig}
\usepackage{bbold}
\usepackage{amssymb}
\usepackage{amsbsy}
\usepackage{color}
\usepackage{slashed}
\usepackage{afterpage}
\usepackage{psfrag}
\usepackage[noabbrev]{cleveref}
\usepackage{dsfont}
\usepackage{enumerate}
\usepackage[shortlabels]{enumitem}
\usepackage[utf8]{inputenc}
\usepackage{amsmath}
\usepackage{graphicx}
\usepackage{graphbox}
\usepackage{braket}
\usepackage{soul}
\usepackage{multirow}
\usepackage{hhline}
\usepackage{abbr}
\usepackage{comment}
\usepackage[normalem]{ulem}

\newcommand{\nnn}{\newline\newline\noindent}
\newcommand{\bstheta}{\boldsymbol{\theta}}

\newcommand{\as}{\alpha_\star}
\newcommand{\fs}{f_\star}
\newcommand{\fst}{f_{\star, 10}}
\newcommand{\ts}{t_\star}
\newcommand{\Mt}{M_{\rm turn}}

\newcommand{\LXu}{{\rm erg \,yr/(s} \, M_\odot {\rm)}}
\newcommand{\mwdm}{m_{\rm WDM}}
 \newcommand{\Mcut}{M_{\rm WDM}^{\rm
  cut}}
\newcommand{\epsrel}{\varepsilon_{\rm rel}}

\definecolor{light-gray}{gray}{0.9}
\definecolor{dark-green}{rgb}{0.0, 0.5, 0.0}
\definecolor{myyellow}{rgb}{1.0, 0.65, 0.0}
\definecolor{myblue}{rgb}{0.0, 0.48, 0.65}

\title{\boldmath 
Simulation-based inference on warm dark matter from HERA forecasts 
}

\author[a]{Q. Decant,}
\author[b]{A. Dimitriou,}
\author[a,c]{L. Lopez-Honorez,}
\author[b]{B. Zaldivar,}

\affiliation[a]{Service de Physique Th\'eorique, C.P. 225, Universit\'e Libre de Bruxelles,\\ Boulevard du Triomphe, B-1050 Brussels, Belgium}
\affiliation[b]{Instituto de Física Corpuscular (IFIC), Consejo Superior de Investigaciones Científicas
(CSIC) and Universitat de València 46980, Valencia, Spain}
\affiliation[c]{Theoretische Natuurkunde \& The International Solvay Institutes, \\ Vrije Universiteit Brussel, Pleinlaan 2, B-1050 Brussels, Belgium}

\emailAdd{quentin.decant@ulb.be}
\emailAdd{llopezho@ulb.be}
\emailAdd{androniki.dimitriou@ific.uv.es}
\emailAdd{bryan.zaldivar@ific.uv.es}

\abstract{ The redshifted 21cm signal from Cosmic Dawn promises to open a new window into the early history of our universe and enable the probing of an unprecedented comoving survey volume. In this work, we revisit the imprint of Warm Dark Matter (WDM) on the 21cm signal power spectrum using an updated implementation of the WDM effect in the public code {\tt 21cmFast} and considering a single population of cosmic dawn galaxies. By focusing on inferring the WDM mass, we analyze the degeneracies between the latter and the astrophysics parameters characterizing star formation and X-ray heating and we emphasize 
the role of the threshold mass for star-forming galaxies, $M_{\rm turn}$. We study the capability of the recently built HERA telescope to reconstruct the WDM mass by adopting the statistical approach of simulation-based inference. We include a comparison of the per-parameter reconstruction quality for different number of simulations used in the training of the algorithm. Our results indicate that HERA could surpass current Lyman-$\alpha$ forest constraints if Cosmic Dawn galaxies exhibit a threshold mass  $M_{\rm turn}\lesssim  10^{8}\, M_\odot$. The X-ray source properties considered in this study may also influence the strength of the WDM constraint for lower threshold masses. 
}


\begin{document}
\maketitle
\flushbottom

\section{Introduction}
\label{sec:introduction}

21cm Cosmology is expected to open a new window on our understanding of the history of the Universe in a redshift range beyond the reach of any existing experiments. The redshifted 21cm signal arising from the so-called Cosmic Dawn (CD) until the Epoch of Reionization (EoR)  shall soon be probed to a high signal-to-noise ratio thanks to ground-based radio interferometers such as the Hydrogen Epoch of Reionization Array (HERA)~\cite{HERA:2022wmy} and the Square Kilometer Array (SKA) experiment ~\cite{Mellema13, Koopmans:2015sua}. The former, though, is already fully built and running, having produced its first results in 2021 \cite{HERA:2021bsv}. An interesting possibility is to study the dark matter hypothesis using the 21cm signal. In this work, we focus on the HERA probes of thermal warm dark matter (WDM) candidates, that suppress small scale structures due to WDM free-streaming, as a benchmark scenario for non-cold dark matter (NCDM) models~\cite{Archidiacono:2019wdp, Hooper:2022byl, Schneider:2018xba}. These DM models can leave a measurable imprint on the 21cm signal delaying the appearance of peaks and troughs both in the sky-averaged signal and in the power spectrum of 21 cm fluctuations, see e.g. \cite{Sitwell:2013fpa, Escudero:2018thh, Schaeffer:2023rsy,Safarzadeh:2018hhg, Lidz:2018fqo,Lopez-Honorez:2018ipk,  Munoz:2020mue,Jones:2021mrs,Giri:2022nxq, Hibbard:2022sng, Schosser:2024aic,Verwohlt:2024efh,Flitter:2023mjj}.

Modeling the evolution of the 21cm signal from EoR and CD is challenging due to the complex interplay of many astrophysical processes, as well as potential physics beyond the Standard Model. 
Consequently, the simulations are computationally costly, even those based on semi-numerical methods as performed by the {\tt 21cmFast} public code \cite{Mesinger:2010ne, Murray:2020trn}, which we have adapted for the present analysis. Clearly, the complexity of the simulations hinders the corresponding statistical analyses, typically relying on likelihood-based approaches such as Fisher matrix analysis or  Markov Chain Monte Carlo (MCMC), see e.g.~\cite{Giri:2022nxq, Hibbard:2022sng, Verwohlt:2024efh, Flitter:2023mjj} for NCDM-related studies. In our analysis, we use a likelihood-free approach instead, dubbed Simulation-based Inference (SBI), which has proven to computationally surpass standard MCMC methods
for an increasing number of problems in areas such as cosmology \cite{mancini2024,List:2023jwo,modi2023,2024JCAP...09..032D}, astrophysics, and particle physics\footnote{Given the high number of works in these areas, we have decided to cite here only the most recent ones.}.   

SBI is already being used in studies concerning 21cm Cosmology, see e.g.~\cite{Zhao:2022ren, Saxena:2023tue, Prelogovic:2023uww,Prelogovic:2024ips, Greig:2024pdm,Schosser:2024aic,Greig:2024zso}. The works~\cite{Zhao:2022ren, Saxena:2023tue, Prelogovic:2023uww,Prelogovic:2024ips, Greig:2024pdm,Greig:2024zso} focus on the ability of SKA to reconstruct astrophysics parameters related to reionization and star formation, while considering the cold dark matter (CDM) scenario, unlike the present study which focuses on WDM instead. In particular,~\cite{Zhao:2022ren} focus as well on the HERA experiment and considers an SBI flavour (``DELFI'') designed to estimate the likelihood function itself. In~\cite{Saxena:2023tue, Greig:2024pdm} the SBI flavour used in the analysis is the same as the one used here, called Neural Ratio Estimation (NRE, cfr. Sec.~\ref{sec:SBI}). 
In \cite{Prelogovic:2024ips}, they compare different summary statistics of the signal in terms of the Fisher Information, while reporting their results based on a combination of fiducial values of the parameters, given that there are no physics-motivated values a priori. Here instead, we report our results for a selection of fiducial scenarios. 

On the other hand, \cite{Schosser:2024aic} does analyze a WDM scenario aiming at reconstructing the WDM mass. They do so with a different SBI method, so-called Neural Posterior Estimation (NPE), in the context of the SKA experiment. Note that the galaxy parametrization used in that work corresponds to an older implementation in {\tt 21cmFast} than the one used here. Furthermore, the default WDM implementation in {\tt 21cmFast} is based  on~\cite{Barkana:2001gr} and makes use of a top-hat window function in the halo mass function (HMF), complemented with a sharp cut in the integrated number of ionizing photons, X-ray emissivity, and Lyman-$\alpha$ photon flux at a certain WDM Jeans mass computed from the WDM free-streaming length. Note that the top-hat window function is expected to give rise to a  significant over-estimation of the number of low-mass haloes in a WDM cosmology, see~\cite{Schneider:2013ria, Benson:2012su},  while those haloes play a key role in estimating the WDM imprint on the 21cm signal. Here, in contrast,   we modified {\tt 21cmFast} to use a sharp-k window function to evaluate the HMF. The latter has been shown to provide a better fit to numerical simulations up to redshifts $z \sim 20$, see~\cite{Schneider:2018xba} (and also~\cite{Lopez-Honorez:2017csg} for a discussion).   Furthermore, beyond a different implementation of the SFR and of the WDM HMF, the threshold mass for star formation, $M_{\rm turn }$,   is effectively kept in \cite{Schosser:2024aic} within the upper end of the $M_{\rm turn }$ range considered here.\footnote{Our $\Mt$ is equivalent  to the $T_{\rm vir}$ (the minimum virial temperature parameter) parameter used in \cite{Schosser:2024aic} but on a smaller range of values. } As we will show, this is expected to hinder the sensitivity to the WDM imprint. 
Other relevant works (based on Fisher matrix or MCMC instead of SBI) on  WDM or related non-cold dark matter scenarios, using the brightness temperature fluctuations in the 21cm signal as a probe, have already been provided; see e.g.~\cite{Giri:2022nxq, Verwohlt:2024efh,Hibbard:2022sng,Jones:2021mrs}. Most recently, 
the analysis of~\cite{Giri:2022nxq}, as well as in~\cite{Verwohlt:2024efh}, were performed using other simulation codes than {\tt 21cmFast}. Those codes in particular allowed to perform MCMC analysis making use of the default assumption of a Gaussian likelihood. 

In this work, we reassess the WDM imprint on the 21cm signal, as potentially measured by the fully deployed HERA experiment, starting with a description of the assumed WDM implementation. We first provide an MCMC analysis of the viable WDM scenario parameter space that is in agreement with existing UV luminosity data, and then analyze the expected degeneracies of the WDM imprint with the astrophysics parameters controlling the overall shape of the 21cm power spectrum. One of our main goals is the assessment of the role of the threshold mass for star-forming galaxies, $M_{\rm turn}$,
in probing thermal WDM. Note that this threshold value is intuitively expected to affect the
21cm signal in the same way as WDM, since it effectively prevents the smallest haloes from contributing to the signal. We evaluated under which conditions HERA is expected to surpass the existing sensitivity of the Lyman-$\alpha$ forest data to the WDM mass, see~\cite{Palanque-Delabrouille:2019iyz}.\footnote{Other recent constraints on WDM from other probes can also be found in e.g.~\cite{Keeley:2024brx,Tan:2024cek} and references there-in. Here we use the exclusion of a 5.3 keV WDM mass at 95\%CL from ~\cite{Palanque-Delabrouille:2019iyz} as a reference.}
We also carefully report on the performance of our SBI configuration to reconstruct the astrophysics parameters.

 In our analysis, we adopt a $\Lambda$CDM cosmology with parameters ($\Omega_{\mathrm{m}}, \Omega_{\mathrm{b}}, \Omega_{\mathrm{\Lambda}}, h, \sigma_8$, $n_s$) = (0.31, 0.049, 0.69, 0.68, 0.81, 0.97) following the {\it TT,TE,EE +lowE +lensing +BAO} result of Planck 2018 \cite{Planck2020A&A...641A...6P}. 
For modelling the 21cm signal during the CD-EoR period, specifically its power spectrum (PS) summary statistics, we use our own modified version of the {\tt 21cmFast} public code~\cite{Mesinger:2010ne, Murray:2020trn}, with the flexible galaxy parameterization of~\cite{Park:2018ljd}. We model the thermal noise by making use of the estimates from {\tt 21cmSense}~\cite{2013AJ....145...65P,2014ApJ...782...66P}. We consider a seven-dimensional parameter space, including the WDM mass, and six other astrophysics parameters expected to show some level of degeneracy with the former. For simplicity, and to minimize computational expense, the astrophysics
model considered here involves a single population of galaxies.   A similar approach was recently employed in an SBI analysis within the context of 21cm cosmology in CDM scenarios to assess the impact of the HMF parametrization on the reconstruction of the astrophysics parameters, see~\cite{Greig:2024zso}.
In our work,  we vary the threshold mass  $M_{\rm turn}$  between  $10^5$ and $10^{10}$ $M_\odot$. Masses larger than $10^8 M_\odot$ at $z=10$ could be representative of atomic-cooling galaxies (ACGs) that we have observed at late times~\cite{Barkana:2000fd}. The ACGs are expected to be dominated by Pop II stars. On the other hand, the lower end of our threshold mass range would include the very first galaxies, hosted by so-called minihalos (with a virial mass $M \lesssim 10^8{\rm M}_\odot$ at $z\sim 10$) and referred to as molecular-cooling galaxies (MCGs). They are expected to predominantly host Population III stars, whose detailed properties are still open to debate. For a discussion on this topic, see e.g.~\cite{Barkana:2001gr,Qin:2020xrg}. The single population of galaxies approach followed here serves as a simple set-up to evaluate the sensitivity of 21cm Cosmology to a modification of the small-scale power spectrum and the halo mass function. It will highlight the important effect of the galaxy population assumption on the 21cm sensitivity reach.

 We organize this paper as follows. In Sec.~\ref{sec:21cm_physics}, we briefly summarize the parametrization of the astrophysics and WDM imprint on the 21cm signal considered here, as well as the expected degeneracies. In Sec.~\ref{sec:analysis}, we provide the details of our SBI analysis, present our results in Sec.~\ref{sec:results}, and  conclude in Sec.~\ref{sec:conclusion}.


\section{21cm Cosmology and astrophysics model}
\label{sec:21cm_physics}
We start with a brief review of 21cm Cosmology and the astrophysics parameters that are expected to have a noticeable impact on the inference of the warm dark matter mass from the power spectrum.

\subsection{The redshifted 21cm signal}
\label{sec:21signal}

The redshifted  cosmic 21cm signal, arising from the hyperfine spin-flip transition of neutral hydrogen at different redshifts, is characterized by the differential   brightness temperature of the 21cm signal with respect to a radio background source, which can be approximated by~\cite{Furlanetto2006PhR...433..181F}
\begin{equation}
		\delta T_\mathrm{b} \approx 20\mathrm{mK} \left(1-\frac{\mathrm{T_\mathrm{CMB}}}{T_\mathrm{S}}\right) x_{\mathrm{HI}} (1+\delta_{\rm b}) \left(1+\frac{1}{H}\,\frac{\mathrm{d}v_r}{\mathrm{d}r}\right)^{-1}\sqrt{\frac{1+z}{10}\frac{0.15}{\Omega_\mathrm{m}h^2}}\frac{\Omega_\mathrm{b}h^2}{0.023}\,,
 \label{eq:deltaTb}
\end{equation}
when considering the Cosmic Microwave Background (CMB), with temperature $T_{\rm CMB}$, as the radio background. In eq.~(\ref{eq:deltaTb}),  $x_{\rm HI}$ is the neutral hydrogen fraction, $\delta_b$ is the baryon number density contrast, $\Omega_{\rm m,b}$ are the matter and baryon energy densities relative to the critical energy density today, $H$ is the Hubble rate and  $\frac{\mathrm{d}v_r}{\mathrm{d}r}$ is the gradient of the proper velocity along the line of
sight. The 21cm signal is seen in emission or absorption when the CMB and spin temperature, $T_{\rm S}$, differ. The spin temperature quantifies the relative occupancy of the two hyperfine levels of the ground state of neutral atomic hydrogen. For redshifts $z$ below $\sim 30$, it is obtained from the equilibrium balance of absorption/emission of 21cm photons from/to the CMB background and resonant scattering of Lyman-$\alpha$ photons coupling the spin temperature to the gas temperature $T_{\rm k}$. Spatial variations of IGM properties lead to fluctuations in the 21cm signal. In what follows, we refer to the 21cm global signal, $\overline{ \delta T_b}$, as the sky-averaged brightness temperature and to the 21cm power spectrum in terms of the
dimension-full quantity $\overline{\delta T_b^2} \Delta_{21}^2$. The latter is defined as:
\begin{equation}
    \overline{\delta T_b^2} \Delta_{21}^2(k,z)=\overline{\delta T_b^2(z)} \times \frac{k^3}{2\pi^2} P_{21}(k,z)
    \label{eq:PS}
\end{equation}
with  the two-point correlation function corresponding to
\begin{equation}
    \langle  \tilde\delta_{21} ({\bf k},z) \tilde\delta_{21} ({\bf k'},z) \rangle= (2\pi)^3 \delta^D({\bf k}- {\bf k'}) P_{21}(k,z)\,,
\end{equation}
with $\langle \cdot \rangle$, the ensemble average, $\bf k$ the comoving wave vector, and $\tilde\delta_{21} ({\bf k},z)$  the Fourier transform of $\delta_{21} ({\bf x},z)= {\delta T_b}({\bf x},z)/\overline{\delta T_b}(z) - 1$, where $\bf x$ denotes the position vector. 

Specific features can be observed in the redshift evolution of the global 21cm signal and the power spectra at a fixed scale. In particular, the global signal is expected to feature an absorption ($\overline{\delta T_b^2}<0$) trough between the epoch of Lyman-$\alpha$ or Wouthuysen–Field (WF) coupling and the subsequent epoch of heating (EoH);  while the signal shall be seen in emission   ($\overline{\delta T_b^2}>0$) between the EoH and the EoR. The HERA telescope, on which we will focus here, is expected to probe the 21cm power spectrum.  Depending on the scale considered, the latter can feature up to 3 peaks, see e.g. Fig.~\ref{fig:PSnonoise}, corresponding to dominant contributions of the auto-power spectrum of  fluctuations in the Lyman-alpha coupling, in the gas temperature, and in the neutral fraction during the WF, EoH and EoR periods, respectively. The detailed shape of the power spectrum, and its testability, will depend on the IGM temperature and ionized fraction.
These quantities, in turn, are influenced by the radiation emitted by stars in the late Universe. 
In the next section, we briefly introduce the modelisation of the astrophysics sources considered in this work, following its implementation in the public code {\tt 21cmFAST} from~\cite{Park:2018ljd}. Concerning the testability, we focus on the HERA experiment, as it has completed deployment~\cite{HERA:2021bsv}, and is currently analysing data from an extended observational campaign.  A first observational result performed with $71$ antennas (out of the total 331) and only 94 nights of measurement has already provided the most constraining upper bounds on the 21 cm power spectrum at redshifts $z=8$ and 10~\cite{HERA:2022wmy}.

\subsection{Astrophysics parameters}
\label{sec:astro}

\renewcommand{\arraystretch}{1.15}
\begin{table}[t]
	\centering
	\begin{tabular}{ | c | c |  c | c |  }
		\hline
		parameter &  units & range & scale\\
		\hline \hline
                $\Mt$& $M_\odot$   &$[10^5,10^{10}]$ & log\\   
                $\fst$& - &  $[10^{-3},1]$&log\\
                $\as$& -&[0.0,1.0]&  linear \\%
                $\ts$ & - & $[0.1,1.0]$ &linear\\
                $L_X$& erg yr/(s $M_\odot$) &$[10^{38},10^{42}]$  &log \\
                $E_0$&  keV&[0.2,1.5] & linear \\
                \hline
	        $\mwdm$& keV & [3,30]& linear\\
		\hline
	\end{tabular}
	\caption{Type (4th column) and prior ranges (3rd column) of the six astrophysics parameters (1st column) considered in our analysis, together with the DM mass ($\mwdm$). See the text for the definition of the parameters.}
	\label{tab:params}
\end{table}

In this paper, we focus on the astrophysics parameters that are expected to influence the inference of the WDM properties. In what follows, we provide a short description of those parameters.   
We consider a single population of galaxies and follow the astrophysics model implemented in {\tt 21cmFAST} as described in~\cite{Park:2018ljd}. The parameters involved in our analysis, see Sec.~\ref{sec:analysis},  are listed in Tab.~\ref{tab:params} together with the scale (linear or logarithmic) and the ranges of the prior values that have been explored in this work.\footnote{  The selection of parameters is similar to the one that was made in the MCMC analysis of \cite{Giri:2022nxq}, yet with a different astrophysics model and 21cm signal simulator. } 

 The stellar mass of a galaxy, denoted as  $M_\star$, is assumed to be related to the virial mass of the host halo, $M$, through the relation $ M_\star = \fs  M $, where $\fs$ is the stellar-to-halo mass ratio. The latter is assumed to scale as: 
\begin{equation} 
    f_\star = \frac{\Omega_{\rm b}}{\Omega_{\rm m}} \min\left\{1, f_{\star,10}\left(\frac{M}{10^{10}~{\rm M}_\odot}\right)^{\alpha_\star}\right\}  \,,
    \label{eq:fstar}
\end{equation}
where $f_{\star,10}$ and $\alpha_{\star}$  are free parameters in our study. This power-law relation between $f_\star$ and $M$ has been inferred in~\cite{Stefanon2021,Shuntov2022A&A...664A..61S}. 
On the other hand, the star formation rate (SFR)  inside a halo of virial mass $M$ is parameterised by \cite{Qin2020MNRAS.495..123Q}
\begin{equation}
    \dot M_\star = \frac{M_\star }{\ts H(z)^{-1}}
    \label{eq:SFR}
\end{equation}
where  $t_{\star} H^{-1}$ is a characteristic star-formation time-scale. Note that in the parametrization of~\cite{Park:2018ljd}, $\dot M_\star$ appears in the computation of the X-ray luminosity and of the estimation of the non-ionizing UV photon flux, while $M_\star$ drives the number of ionizing photons per baryon. This allows for disentangling the effects of $t_\star$ and $\fst$ on the 21cm signal.

On the other hand, the number density of galaxies per unit halo mass that contributes to the number of UV ionizing and non-ionizing photons, the X-ray emissivity and the Lyman-$\alpha$ flux is accounted for by the halo mass function (HMF), ${{\rm d}n}/{{\rm d}M}$, weighted by a duty cycle, $f_{\rm duty}$. Here we adopt the Sheth, Mo and Tormen (SMT)~\cite{SMT2001MNRAS.323....1S} first crossing distribution in the HMF, and we make use of a sharp-k window function to evaluate the root-mean-square (rms) variance of density perturbations, see more details in Sec.~\ref{sec:TKdndM}. The duty cycle accounts for inefficient star formation in low-mass galaxies due to inefficient cooling and/or feedback mechanisms. It is assumed  to take the following form~\cite{Xu2016ApJ...823..140X, Sobacchi:2013wv, Sobacchi:2013ww, Oshea2015ApJ...807L..12O, Qin2020MNRAS.495..123Q}: 
\begin{equation}
     f_{\rm duty}=\exp\left(-\frac{M_{\rm turn}}{M}\right).
     \label{eq:fduty}
 \end{equation} 
  While e.g.~\cite{Park:2018ljd} concluded that  $M_{\rm turn}< 2\times 10^9\, M_\odot$ to account for UV luminosity data, molecular cooling would point to threshold masses as low as $M_{\rm turn}=10^5\, M_\odot$~\cite{Barkana:2001gr,Qin:2020xrg}.
  In our analysis, we consider a single population of galaxies whose threshold mass is fixed to a constant value $M_{\rm turn}$ lying in the range between $10^5$ and $10^{10}\, M_\odot$.\footnote{ The threshold mass $\Mt$ depends on the minimum virial mass from which the gas can cool, and on feedback processes. Atomic cooling is expected to be efficient above minimum virial temperatures of $T_{\rm vir}\sim 10^4$K, while molecular cooling provides cooling from $T_{\rm vir}\sim 300$K \cite{BARKANA2001125}. At $z=20$, making use of the virial theorem and neglecting feedback processes, this would   correspond to minimal threshold masses of the order of $ M_{\rm mol}= 10^{5}\,M_\odot$  and 
  $ M_{\rm atom}= 2\times 10^7 \,M_\odot$, respectively. See also e.g.~\cite{Sobacchi2013MNRAS.432.3340S,Qin2020MNRAS.495..123Q,Munoz2022MNRAS.511.3657M,Park:2018ljd} for a discussion on the impact of those feedback mechanisms on the threshold mass. An implementation of these effects is provided in {\tt 21cmFAST}, but running simulations taking into account these effects entails a significant computational cost.    } 
  
Finally, the X-ray luminosity, which brings the global signal in emission wrt the CMB, is normalized with the integrated soft-band ($E<2$ keV) luminosity per SFR (in units of ${\rm erg\ yr\ s^{-1}\ M_\odot^{-1}}$), that we denote:
\begin{equation}
L_X= \int^ {2 {\rm keV}}_{E_0}
{\rm d} E \, {{\cal L}_X}\,,
  \label{eq: LX}
\end{equation}
where ${\cal L}_X$ refers to the specific X-ray luminosity per unit star formation rate, and  $E_0$ is the X-ray energy threshold below which photons cannot escape the host galaxy. $L_X$ and $E_0$ are considered as free parameters in our analysis. The X-ray threshold $E_0$ depends on the interstellar medium as well as the environment of the X-ray sources.  Furthermore, the first HERA results~\cite{HERA:2021noe}, analyzed in the context of a CDM cosmology, tend to favor $L_X\geq 10^{40}$ erg yr/(s $M_\odot$). Here, in the context of our WDM analysis, we consider the ranges reported in Tab.~\ref{tab:params}.

\subsection{Warm dark matter}
\label{sec:TKdndM}
Thermal WDM is assumed to be in kinetic and chemical equilibrium with the rest of the plasma before its decoupling, which settles its relic abundance through a relativistic freeze-out. As WDM decouples while being relativistic, it free-streams over a characteristic length scale, referred to as the free-streaming length, that is larger for smaller WDM mass. 
This induces a suppressed linear matter power spectrum at small scales, as well as a reduced number of low-mass halos contributing to astrophysics radiation sources. In practice, WDM implies a delay in the EoR, EoH and WF eras, which gets imprinted in the 21cm signal and related observables, see e.g.~\cite{Sitwell:2013fpa, Lopez-Honorez:2017csg,Escudero:2018thh, Safarzadeh:2018hhg, Lidz:2018fqo,Lopez-Honorez:2018ipk,  Munoz:2020mue,Jones:2021mrs, Schaeffer:2023rsy,Giri:2022nxq,Schosser:2024aic,Verwohlt:2024efh}.
The lower the WDM mass, $\mwdm$, the larger the WDM free-streaming scale, and as a result, the more suppressed the number of low-mass haloes. In Tab.~\ref{tab:params}, the prior on the DM mass is chosen such that for its high-mass end, 30 keV, the DM scenario is equivalent to CDM in the context of our analysis. On the other hand, the low-mass bound of 3 keV is chosen to be such that the WDM candidate is marginally allowed by existing Lyman-$\alpha$ forest bounds on WDM, see \cite{Viel:2005qj,Palanque-Delabrouille:2019iyz}.

 The WDM suppression of the linear matter power spectrum, $P(k)$,  has been parameterized in terms of a transfer function, $T_{\rm WDM}(k)$, which is defined as a function of the ratio of the WDM and the CDM power spectra:   $T^2_{\rm WDM}(k)=P_{\rm WDM}(k) / P_{\rm CDM}(k)$. Assuming that WDM makes up all DM, the transfer function has been fitted to~\cite{Bode:2000gq, Viel:2005qj}:
\begin{equation}
  T_{\rm WDM}(k) = (1+ (\alpha_{\rm WDM} k)^{2\nu})^{-5/\nu} ~,
\label{eq:TWDM}
\end{equation}
where $k$ denotes the amplitude of the wave vector, Fourier dual of a comoving distance. 
In this work, we use $\nu=1.12$ and a breaking scale
\begin{equation}
  \alpha_{\rm WDM}= 0.0445 \left(\frac{{\rm keV}}{m_{\rm WDM}}\right)^{1.11}\left(\frac{\Omega_{\rm WDM}}{0.25}\right)^{0.11}\left(\frac{h}{0.7}\right)^{1.22} \, {\rm Mpc}/h ~,
  \label{eq:alphWDM}
\end{equation}
which indicates the typical scale from which the WDM power spectrum gets significantly suppressed and effectively provides an estimate of the WDM free-streaming length. Notice that in this analysis, we have set the prefactor to 0.0445 ${\rm Mpc}/h$ to account for the case of $m_{\rm WDM}=5.3$ keV, which saturates the current Lyman-$\alpha$ data lower bound~\cite{Palanque-Delabrouille:2019iyz}.   Indeed, as already emphasized in e.g.~\cite{Decant:2021mhj}, previous fits from~\cite{Bode:2000gq, Viel:2005qj} were originally designed for lower mass dark matter.

To simulate the 21cm signal, one has to evaluate the imprint of WDM on non-linearly evolved matter perturbations that have given rise to a distribution of haloes. The halo mass function, i.e., the number of halos per unit mass as a function of mass and redshift, is defined as~\cite{Jenkins:2000bv}
\begin{equation}
\label{eq:dndM}
  \frac{dn(M,z)}{dM} = \frac{\rho_{{\rm m}}}{M^2} \, \frac{d\ln \sigma^{-1}}{d\ln M} \, f(\sigma) ~, 
\end{equation}
where $\rho_{{\rm m}} = \Omega_{{\rm m}} \, \rho_{c}$ is today's matter energy density with $\rho_{c}$ the critical energy density, $\sigma^2=\sigma^2(M,z)$ is the variance of density perturbations on a mass-scale $M$ at redshift $z$, and $f (\sigma)$ is the so-called first crossing distribution. Here we consider the Sheth-Thormen description  given by~\cite{Sheth:1999mn}
\begin{equation}
\label{eq:ST}
f(\sigma) = A \, \sqrt{\frac{2 \, q}{\pi}} \, \left(1 + \left(\frac{\sigma^2}{q \, \delta_{\rm c}^2} \right)^{p} \right)\left(\frac{\delta_{\rm c}}{\sigma}\right) \, e^{-\frac{q \, \delta_{\rm c}^2}{2 \, \sigma^2}} ~\,,
\end{equation}
where $\delta_{\rm c} = 1.686$ is the critical overdensity. 
In this work, we use $p=0.3$, $q=1$, and $A=0.322$~\cite{Schneider:2014rda}. 
The root-mean-square (rms) variance of density perturbations at a given radius $R$, is evaluated using
\begin{equation}
\label{eq:sig2}
\sigma^2(M(R), z) = \left(\frac{D(z)}{D(0)}\right)^{2} \int \frac{d^3k}{(2\pi)^3} \, P_{\rm WDM}(k) \, |W(kR)|^2 ~, 
\end{equation}
where the redshift dependence is driven by the linear growth function,
$D(z)$ and  $P_{\rm WDM}(k)$ is the linear power spectrum at $z=0$ obtained with the transfer function of eq.~(\ref{eq:TWDM}). For this work,  we implemented the Fourier transform of the filter function,  $W(kR)$,  referred to as a sharp-k
window:
\begin{equation}
  \label{eq:SK}
  W(kR)= \Theta(1 - kR) ~, 
\end{equation}
where $\Theta$ denotes the Heaviside function.  The latter has  indeed been argued to account for  the observed WDM cosmology behaviour in N-body simulations, namely
the WDM suppression of the halo mass function at low halo masses~\cite{Benson:2012su,
  Schneider:2013ria}, see also e.g.~\cite{Schneider:2018xba,Hibbard:2022sng} in the context of 21cm cosmology.

\begin{figure}
\centering
\includegraphics[width=0.65\linewidth]{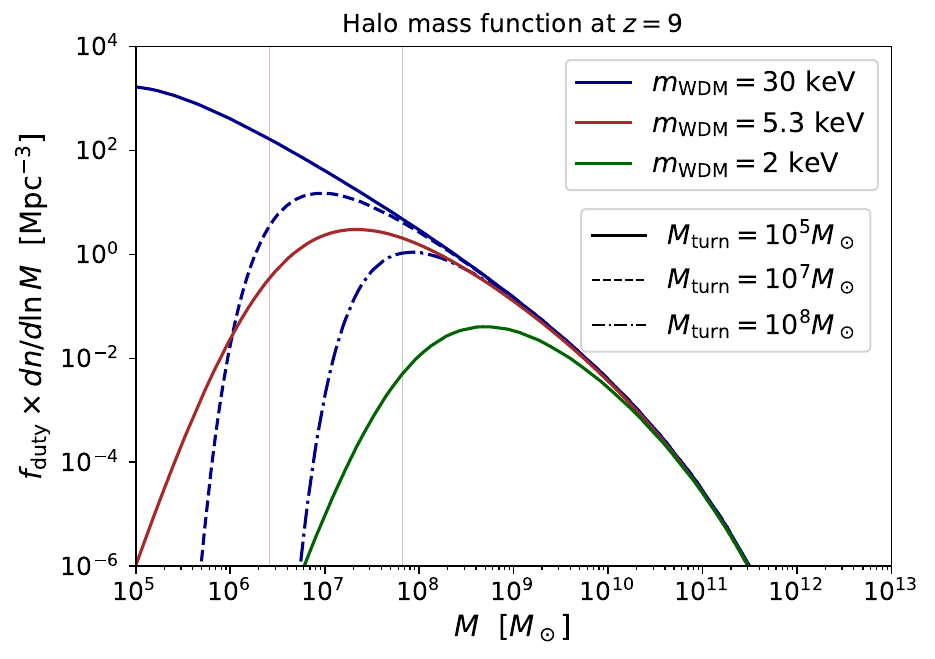}
\caption{Halo mass function as a function of the halo mass at $z=9$.
  With the continuous lines, we show the case of $m_{\rm WDM}= 2, 5.3$,
  and 30 keV WDM mass with green, red and blue colours, assuming $\Mt=10^5\, M_\odot$. The  $m_{\rm WDM}=30$ keV case is shown as a
  representative of a CDM scenario in the context of our analysis. The vertical thin coloured lines indicate the cutoff mass scale,  $M_{\rm
    WDM}^{\rm cut }$, see eq.~(\ref{eq:McutWDM}). This scale provides a rule-of-thumb estimate of the threshold below which halo formation is significantly suppressed due to the free-streaming of WDM particles.  The
  dashed and dot-dashed blue lines correspond to the 30 keV DM HMF
  with a threshold mass increased to $M_{\rm
    turn}=10^7$ and $10^8 \, M_\odot$, respectively.    }
\label{fig:HMF}
\end{figure}

When making use of the sharp-k  (SK) filter, a free parameter $c_{\rm SK}$ has to be
introduced to relate the halo mass to its radius as follows:
\begin{equation}
    M_{\rm SK}(R)= \frac{4\pi}{3} \, (c_{\rm SK} R)^3\,.
\label{eq:MSK}
\end{equation}
This parameter has been set to $c_{\rm SK}=2.5$ in our work, see~\cite{Benson:2012su, Schneider:2014rda,Schneider:2018xba} for references. Considering a radius of the order of the breaking scale, $R=\alpha_{\rm WDM}$, in the SK halo mass of eq.~(\ref{eq:MSK}), we get a corresponding halo mass of:
\begin{equation}
M_{\rm WDM}^{\rm cut} (\mwdm)=2.6 \, \left(\frac{5.3{\rm keV}}{m_{\rm WDM}}\right)^{3.33} \, \times 10^6\, M_\odot\,,
    \label{eq:McutWDM}
\end{equation}
below which the HMF is significantly suppressed due to WDM free streaming.

The imprint of WDM on the HMF is illustrated in
Fig.~\ref{fig:HMF} displaying $f_{\rm duty}\times dn/d\ln M$ as a function of the halo
mass $M $ at $z=9$. The continuous blue line corresponds to a 30 keV WDM, which in practice, is equivalent to CDM in the context of our analysis,\footnote{Indeed, for $\mwdm=30$ keV, $M^{\rm
  cut}_{\rm WDM}\sim 10^4\, M_\odot$, with  $M^{\rm
  cut}_{\rm WDM}$ defined in eq.~(\ref{eq:McutWDM}), which is below the minimum
threshold mass for star formation considered in this work.
} while the red and green lines assume a WDM cosmology with $m_{\rm WDM}=5.3$ and 2 keV,
respectively. The lower the DM mass, the longer the free streaming
scale of eq.~(\ref{eq:alphWDM}) and the larger is the halo mass at
which the HMF is suppressed, as expected from eq.~(\ref{eq:McutWDM}). In particular, $M_{\rm WDM}^{\rm
  cut}$ for $\mwdm=5.3$ and 2 keV are  
indicated with thin colored vertical red and green
lines in Fig.~\ref{fig:HMF}. This WDM halo mass cut-off estimate can conveniently be compared
to the threshold mass for star formation $M_{\rm turn}$ entering in the
duty cycle of eq.~(\ref{eq:fduty}). Indeed, as mentioned above, the
ionization, X-rays and Lyman-$\alpha$ fluxes from stars induce the
21 cm signal for a given halo mass is weighted by the product of
$  f_{\rm duty} \times dn/d\ln M$. With the continuous, dashed and dot-dashed blue
lines, we assume increasing  threshold
masses, $M_{\rm turn}=10^5$, $10^7$ and $10^8\, M_{\odot}$, respectively, dictating the form of the duty cycle factor. The plot suggests that a threshold mass $M_{\rm turn}>10^8\, M_{\odot}$ may hinder the straightforward reconstruction of a WDM cosmology characterized by $m_{\rm WDM}=5.3$ keV. In general, when $M_{\rm turn}\gg \Mcut$, the WDM free-streaming effect is expected to
suppress haloes that are too small to host stars. The 21cm signal
shall then not be sensitive to the WDM imprint. Equivalently, if
$\Mcut \gg M_{\rm turn} $, WDM would prevent haloes from forming that would have otherwise significantly contributed to the onset
of the EoR, EoH and WF eras. This would for example be the case of $m_{\rm
  WDM}=2$ keV with $M_{\rm turn}\lesssim 10^7$. Yet,
such a low mass WDM is already excluded by Lyman$-\alpha$ forest observations from~\cite{Palanque-Delabrouille:2019iyz}.

\subsection{WDM imprint on 21cm signal and degeneracies}
\label{sec:WDM21}

\begin{figure}[t!]
\centering
\includegraphics[width=0.9\linewidth]{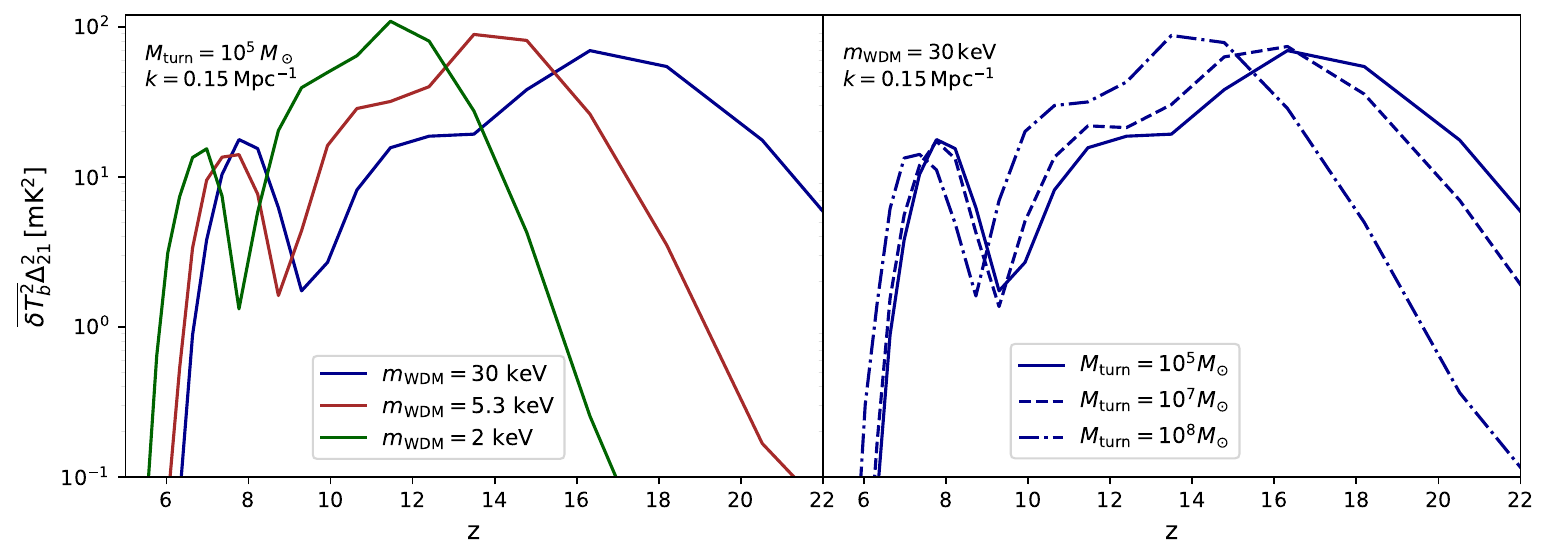}
\caption{  21cm power spectrum   $\overline{\delta T_b^2} \Delta_{21}^2$ as a function of the redshift for fixed scale  $k=0.15$/Mpc. Left: effect of varying the WDM mass. Right: effect of varying the threshold mass for star formation. The astrophysics parameters that are not mentioned in the figure have been fixed to $\log(f_{\star,10})= -0.9$,
     $\alpha_{\star}=0.46$, $E_0=0.5$ keV and $L_X=10^{40}$ $\LXu$.  }
\label{fig:PSnonoise}
\end{figure}

In this section, we briefly comment on the imprint of WDM on the 21cm power spectrum given the parametrization detailed in Sec.~\ref{sec:TKdndM} as well as on the expected degeneracies between $\mwdm$ and the astrophysics parameters introduced in Sec.~\ref{sec:astro}. For that purpose the 21cm power spectrum $\overline{\delta T_b^2} \Delta_{21}^2$ is illustrated in  Fig.~\ref{fig:PSnonoise} as a function of the
 redshift at a fixed scale $k=0.15$
 Mpc$^{-1}$ for a selection of parameter values. The continuous blue lines correspond to a fiducial
 scenario assuming $M_{\rm turn}=10^5\, M_\odot$ with $m_{\rm WDM}=30$
 keV in both plots. In the left plot, we illustrate the effect of the
 WDM mass, the red and green curves corresponding to  a DM mass of $\mwdm= 5.3$
 and 2 keV, respectively.  
 We see that the lower the WDM mass, the later the different features of the 21cm power spectrum appear. Indeed, as discussed in Sec.~\ref{sec:TKdndM}, lower WDM
 masses induce larger DM-free streaming length at the time of structure
 formation, preventing the lowest mass halos from forming efficiently. This
 effectively reduces the ionizing, X-ray, and Lyman-$\alpha $ photons
 luminosity from cosmic dawn galaxies, delaying the EoR, EoH and WF
 eras. On the other hand, in the right plot, we illustrate the impact of
 different choices of threshold mass for star-forming galaxies in a CDM-like cosmology, the
 dashed and dot-dashed blue lines corresponding to $M_{\rm turn}=10^7$
 and $10^8\, M_\odot$, respectively. Increasing $M_{\rm turn}$ prevents the lower mass haloes to host stars. Again, this gives rise to a shift of the power spectrum features to later redshifts.
  A higher value of the threshold
 mass $M_{\rm turn}$ has a qualitatively similar effect on the power spectrum 
 as a lower $\mwdm$. 
  We expect thus a positive correlation between $M_{\rm turn}$
 and $\mwdm$ in our analysis as an increase in $\mwdm$ could potentially be compensated by an increase of $\Mt$. Also, relatively large
 values of $M_{\rm turn}$ shall prevent any sensitivity of the 21 cm
 signal to the  WDM properties beyond minimal $\mwdm$ as the suppressed haloes can not host
 stars. 
The effect on the 21cm power spectrum of the other parameters of Tab.~\ref{tab:params}  is as follows.  The parameters $\as$ and $\fst$ regulate the fraction of stars into haloes, while $\ts$ further tunes the amplitude of the SFR. Increasing
$\alpha_\star$ favours larger haloes in hosting stars, which can
effectively be mimicked with decreasing $\mwdm$, thereby suppressing the number of low mass haloes. We expect
the $\mwdm$ and $\alpha_\star$  parameters to be positively correlated e.g. an increase in $\alpha_\star$ can be compensated by an increase of $\mwdm$. On the other hand,  $\fst$ defines the overall amplitude of the fraction of stars into haloes (independently of the halo masses). Increasing $\fst$ enhances any source of radiation, leading to the emergence of earlier features in the power spectrum.
An increase of $\fst$ in a WDM scenario could thus be mimicked by an increase in $m_{\rm WDM}$. Those parameters could therefore appear to be anticorrelated when the WDM imprint is not screened by a large value of $M_{\rm turn}$. On the other hand, $t_*$ regulates the overall amplitude of the SFR that is relevant for non-ionizing UV and X-ray luminosities. Larger $\ts$ suppresses the SFR, which can be mimicked with lower $\mwdm$. As a result, $\ts$ shall display a positive correlation with $\mwdm$, as is the case of $\Mt$ and $\as$. 
The above degeneracies will be further studied in Sec.~\ref{sec:2Dposteriors}.  

The last two parameters mainly affect the EoH. $L_X$ regulates the overall amplitude of the X-ray emissivity at fixed energy threshold $E_0$. Considering low $E_0$, we make the
X-ray spectrum is softer, i.e. involving X-rays with shorter mean free
path. The latter induces an inhomogeneous heating of the
IGM~\cite{Pacucci:2014wwa} which induces a large EoH peak. On the other hand, e.g. for low $E_0=0.5$ keV, one can reduce the amplitude of the power spectrum at a given redshift by increasing $L_X$ as this might shift the EoH peak to an earlier time, where HERA has no sensitivity, see e.g.~\cite{Greig:2017jdj} for illustration. Changing $E_0$ is thus expected to show some level of  degeneracy with
changing $L_X$ and could indirectly affect the $\mwdm$
reconstruction. This is in particular further illustrated in Sec.~\ref{sec:bounds}.

\section{Analysis}
\label{sec:analysis}

In this section, we describe the different steps of our analysis. We start by imposing constraints on our parameters of interest from UV luminosity data in Sec. \ref{sec:MCMC}. We then describe the simulations of the power spectrum in Sec. \ref{sec:Mock} before introducing our statistical analysis (the SBI implementation) in Sec. \ref{sec:SBI}.

\subsection{Priors from the UV luminosity data with MCMC}
\label{sec:MCMC}

\begin{figure}[t!]
\centering
\includegraphics[width=0.95\linewidth]{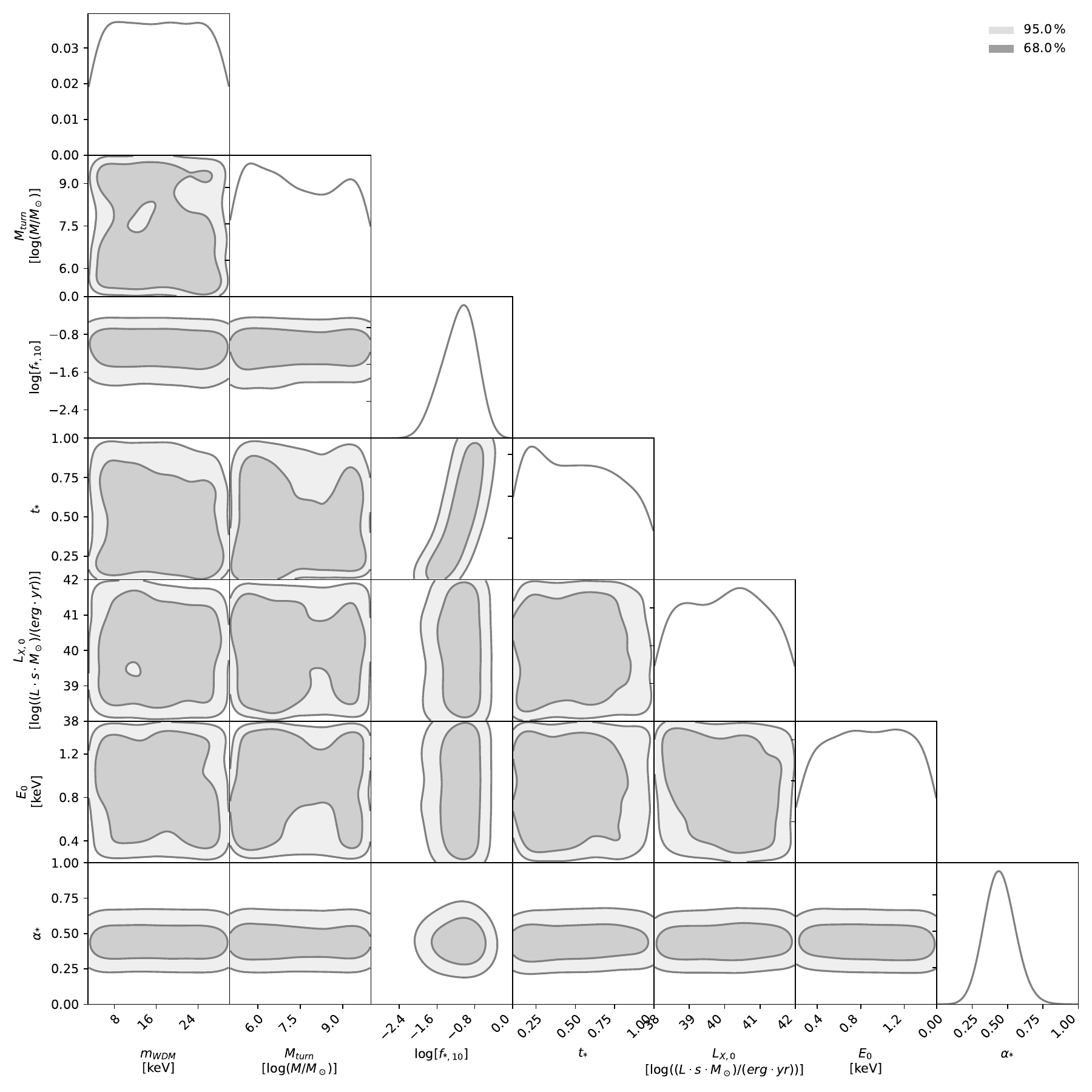}
\caption{ Results of the MCMC analysis taking into account UV-luminosity data constraints. Darker (lighter) contours correspond to 68\% (95\%) credible intervals.  
The range of the axes is made to coincide with the prior ranges of Table \ref{tab:params}.} 
\label{fig:MCMC}
\end{figure}

Some of the astrophysics parameters discussed in Sec.~\ref{sec:astro} can already be constrained by existing UV luminosity data, see e.g.~\cite{Mirocha:stw2412,Park:2018ljd, HERA:2021bsv}.  In contrast to ionizing UV photons, the soft UV responsible for coupling the gas and spin
temperatures can have long mean
free paths, even through the neutral IGM. The non-ionizing UV luminosity functions, $\phi$,  are evaluated with {\tt 21cmFAST} assuming:
\begin{equation}
    \phi (M_{\rm UV})= f_{\rm duty}  \frac{dn}{dM} \times \frac{dM}{dM_{\rm UV}}\,.
    \label{eq:UV}
\end{equation} 
The last factor accounts for the conversion of the halo mass to UV magnitude. It is evaluated assuming that the SFR, $\dot M_\star$ introduced in eq.~(\ref{eq:SFR}), is proportional to the rest-frame UV luminosity of a galaxy which itself is a function of the UV magnitude $M_{\rm UV}$, see~\cite{Park:2018ljd} for details. In this paper, we use the data compiled by~\cite{Bouwens2021AJ} and restrict the analysis to  $M_{\rm UV}> -20$ and $z\geq 6$. 
As in previous works~\cite{HERA:2021bsv, Park:2018ljd,Greig:2024zso},  the latter procedure is done to
 avoid uncertainties related to the estimation of dust extinction, which becomes important at low $z$ and high brightness, see e.g.~\cite{Sabti:2020ser}.    

We compare the data from~\cite{Bouwens2021AJ}  to the UV luminosity functions obtained with  {\tt 21cmFAST} for a given choice of parameters in an MCMC analysis with a Gaussian likelihood. The resulting posterior distributions are shown in Fig.~\ref{fig:MCMC}. Except for $\alpha_\star$ and $\fst$, constrained through the dependence of $\dot M_\star$ in   ${dM}/{dM_{\rm UV}}$, the posteriors do not significantly constrain the range of the parameters considered here. The favoured range of the star formation rate parameters\footnote{Notice that $\log$ refers to $\log10$ in our paper. } are $\log (\fst) \in \{-1.29, -0.77\}$ and $\alpha_\star \in \{0.35, 0.50\}$ at 68\% CL.
In our SBI analysis, we will make use of the posteriors depicted in Fig.~\ref{fig:MCMC} as priors for our parameters. Notice that the UV-luminosity function data considered here give rise to distributions compatible with those obtained  in~\cite{Park:2018ljd} for comparable ranges of the parameter values, even though we consider WDM cosmologies with a range of $M_{\rm turn}$ extended to lower values and a more recent UV luminosity data set.

Note that the optical depth to reionization measurements from Planck are also well known to impose constraints on the astrophysics parameters, in particular, on the parameters driving the fraction of ionizing photons escaping the galaxies. Yet, to estimate the ionized fraction, a lightcone has to be simulated which induces a significant time and computational cost compared to the evaluation of the UV luminosity functions.
Furthermore, a dedicated analysis accounting correctly for the history of reionization resulting from the simulations, and not from a default $\tanh$ reionization assumed for Planck analysis, could be needed, see e.g.~\cite{Qin:2020xrg} for such study in a CDM context.  To avoid such an extra non-negligible cost of computational resources, in our analysis, we fix the escape fraction parameters to the central values obtained in~\cite{HERA:2021noe} from UV + optical depth + HERA I data, namely $\log( f_{\rm esc, 10})=-1.1$ and $\alpha_{\rm esc}=0.02$. In the latter case,    we checked in the course of our study that the optical depth measurement from Planck did not significantly affect the posteriors of the parameters presented in Fig.~\ref{fig:MCMC}. All other parameters, not included in Tab.~\ref{tab:params}, are set to their default values in {\tt 21cmFast}. 

\subsection{21 cm Mock data}
\label{sec:Mock}

For our SBI analysis, we simulate 21 cm signal lightcones with {\tt 21cmFast} within boxes of $[300 \,{\rm Mpc}]^3$ comoving volume on 
a $[200]^3$ cells grid. From those lightcones we obtain 21cm cosmological power spectra on fixed $k$ and $z$ grids reported in the App.~\ref{app:moreMock}, see Tab.~\ref{tab:zk_bins}.  Note that, to ensure the stochasticity of our mock data set, we picked a new {\it random seed} for each realization of the cosmological 21cm power spectrum obtained from {\tt 21cmFast}, see e.g.~\cite{Prelogovic:2023uww}. We add to each simulated cosmological power spectra several realizations of the thermal noise contribution, see the discussion in Sec.~\ref{sec:SBI}.  
The thermal noise is assumed to follow a zero-mean Gaussian distribution with variance given by the experimental sensitivity, $\sigma^2_{\rm therm}(k,z)$, associated with the HERA telescope.  We sample noisy data from such a Gaussian truncated at zero to enforce positive power spectrum values. 
To compute the sensitivities, we use the {\tt 21cmSense} code~\cite{2014ApJ...782...66P,2013AJ....145...65P}.
Within the latter, the HERA
experiment is described as a hexagonal array of 331 antennas of 14~m
dish size. In addition, we have specified a bandwidth of $B=8~{\rm MHz}$, a
spectral resolution of $\delta \nu \sim 97.7~{\rm kHz}$, and
a total operating time of 1000 hrs ($\sim$ 167 nights with 6 hours of
observation per day).  The thermal noise depends on the
system temperature, $T_{\rm sys}$, as $\sigma_{\rm therm}(k, z)\propto T_{\rm
  sys}^2(z)$~\cite{2013AJ....145...65P}, where the frequency dependence is assumed to follow~\cite{DeBoer:2016tnn}:
\begin{equation}
    T_{\rm sys}(z) = 100~{\rm K} + 120~{\rm K}\left(\frac{\nu(z)}{150 ~{\rm MHz}}\right)^{-2.55} \,.
    \label{eq:Tsys}
\end{equation}

\begin{figure}[t!]
\centering
\includegraphics[width=1\linewidth]{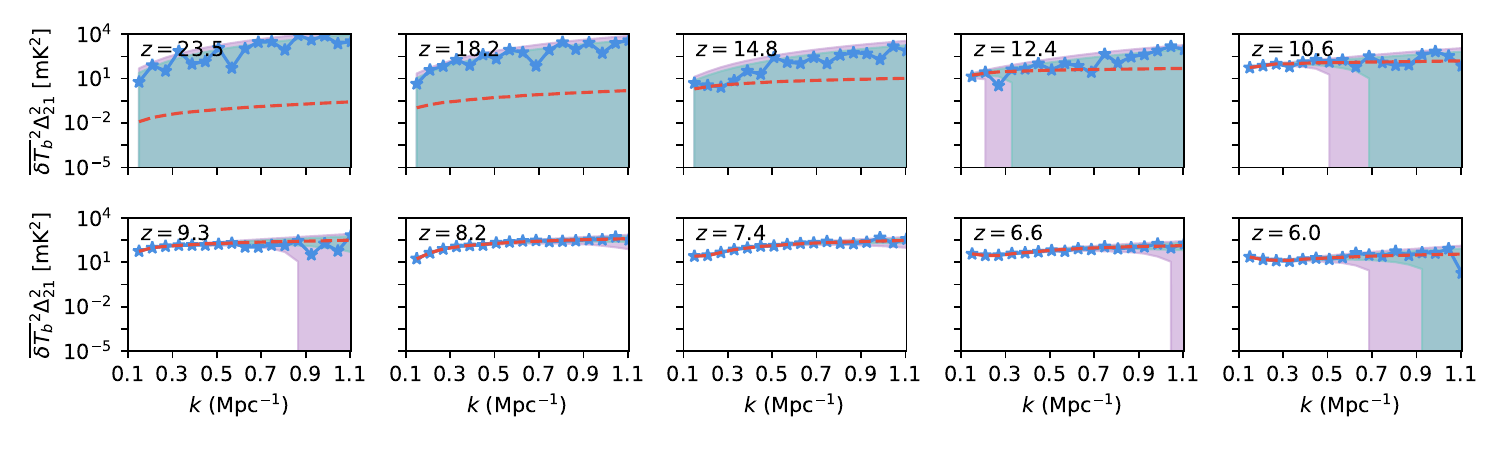}
\caption{  Example of mock 21cm power spectrum $
 \overline{\delta T_{b}}^2 \Delta_{21}^2$ as a function of the wave number $k$ at different redshifts for
  $m_{\rm WDM}= 5.3$ keV, $M_{\rm turn}= 10^{7}~\textup{M}_\odot$. 
  The dashed red lines show the noiseless power spectrum; the blue starred lines are noisy realizations of the power spectrum, while the blue (purple) bands correspond to 68\% (95\%) CL uncertainties. See text for details. The rest of parameters have been fixed as in Fig. \ref{fig:PSnonoise}.}
\label{fig:PSnoise}
\end{figure}

\noindent 
HERA will measure the redshifted 21cm signal in a frequency band of 50-250 MHz, which shall allow to probe the redshift range $z \sim 4.6-27$, with $z=\nu_{21}/\nu-1$ and $\nu_{21}= 1420.4$ MHz is the rest frame frequency of the 21cm signal. The bandwidth $B$ sets the size of frequency bins, which, expressed in terms of redshifts, implies increasing the size of redshift bins when increasing redshift. In this work we use a bandwidth  $B=8$ MHz and 19 redshift bins in redshift range  $[5.9 - 25.3]$, see Table \ref{tab:zk_bins}. The upper bound has been set to 25.3, as larger redshift power spectra get large noise contributions. 
Furthermore,  given the HERA configuration, the accessible $k$ modes are mostly dominated by the $k$  component parallel to the line of sight (l.o.s.), $k_\parallel$\cite{HERA:2021bsv}. Mapping angles in the sky and frequency to cosmological distances, one shall thus consider  $k$-bins set by  $\Delta k_\parallel(z, \Delta \nu) \equiv 2\pi (\nu_{21}/ {\Delta \nu})\times H(z)/(1+z)^2$, where $\Delta \nu$ refers to a frequency step, implying that $\Delta k_\parallel(z, \Delta \nu)$ decreases with increasing redshift. In this analysis, we use fixed values of the $k$ bins width. We set the latter to the the maximal value of $\Delta k_\parallel$, obtained at the minimum redshift, i.e. $\Delta k= \Delta k_\parallel(z_{\rm min}, B)=0.06$ Mpc$^{-1}$. As a result, we use 15 $k$-bins, with our minimum $k$-bin center  set to 0.15 Mpc$^{-1}$, and the maximum $k$-bin center value is set to 0.99 Mpc$^{-1}$ to avoid large thermal noise, see Table \ref{tab:zk_bins}.

 Figure~\ref{fig:PSnoise} illustrates the case of $m_{\rm WDM}=5.3$ keV and $M_{\rm turn}=10^{7} \textup{M}_\odot$ with the other parameters set to the same fiducial values than in Fig.~\ref{fig:PSnonoise}. The red dashed lines depict the noiseless power spectra as a function of the wave mode $k$ for the entire $ k$ range considered in our analysis. Note that a selection of 10 values of the redshifts, out of the 19 considered redshift bins, is displayed.  
The blue and purple areas indicate the  68\% and 95\% CL uncertainty bands around the noiseless power spectra. In particular, the 
  blue lines with stars, indicating the $k$-bin centres, provide one realization of the noisy simulated power spectra. 
  As expected from eq.~\eqref{eq:Tsys}, we see in Fig.~\ref{fig:PSnoise} that the thermal noise becomes more important for the larger redshift values considered, dominating completely the signal for $z\gtrsim 12$ in all $k$ range. On the other hand, for the fiducial values of the parameters considered here, the thermal noise becomes negligible for $6\lesssim z\lesssim 8$,  while it rises again for $z\lesssim 6$. 
Notice that the variation of the relative importance of the thermal noise contribution, as a function of the redshift, depends on the fiducial parameter values considered.

\subsection{Simulation-based inference}
\label{sec:SBI}

The general working principle of SBI is to compare an observed dataset with a sufficiently large set of simulated datasets (provided we have a simulator at hand), each of which has known values of the input parameters. As such, by selecting somehow the simulated datasets closest to the observed one, we can deduce the distribution of values of the parameters behind the latter. Of course, this process entirely relies on the assumption that the simulator captures reasonably well the physics responsible for the phenomena that we are observing.  If this was not the case, the whole procedure would surely give misleading results.  While conceptually the SBI approach is not new,\footnote{As far as we know, likelihood-free methods appeared already in the 1980s, with the Approximate Bayesian Computation (ABC) method \cite{ABC}.} the exponential advance in computing power allowing to work with largely parameterized Machine Learning (ML) models as deep neural networks and others have unlocked the potential of SBI to solve modern problems in science requiring complex simulations. This is e.g. the case in areas like cosmology, astrophysics, particle physics, and others \cite{2023mla..confE..18L,PhysRevD.108.042004,2024PhRvD.109h3008A,2024JCAP...09..032D,2024RASTI...3..724S,2024PhRvD.110h3010P,Prelogovic:2023uww,2023PhRvL.131q1403Z,PhysRevD.109.123547,2024NatAs...8.1457H}. Modern SBI flavors use these ML models to approximate the posterior distribution of the input parameters. 

SBI differs from the MCMC approach in three essential ways. First, it is an approximate inference method that provides samples from approximations of the true posterior. Second,  it is a likelihood-free method, referring to the fact that there is no need to evaluate a likelihood function. Third, it can be designed as a fully ``pre-trained'' strategy, meaning that the inference on new datasets takes essentially no computational time after the model has been fitted/trained to a large number of simulated datasets. Precisely the latter point is behind the computational advantages of SBI over MCMC seen in many analyses. At this point, it is important to note that none of the two approaches can guarantee to give, in practice, unbiased results. Indeed,  while MCMC is surely unbiased provided the true likelihood is available (which is hardly the case for data coming from complex simulators\footnote{Modulo observables for which it makes sense to apply the Central Limit Theorem, thus being Gaussian in practice.}), SBI, on the other hand, in any of its flavours, approximate at the end of the day the posterior distribution in one way or another, thus being inherently biased. 
For sufficiently complex and large parametrisations (as in the case of modern Machine Learning models), such approximations can however become increasingly better with respect to the true unknown posterior distribution.  On the other hand, the scaling of the computational time with the number of parameters entering the simulation has been shown to be much better with SBI as compared to typical flavours of MCMC~\cite{Miller:2020hua}.

As in most cases, the simulator we use in this work, contains, on the one hand, physical parameters of interest $\boldsymbol{\theta}$ and, on the other hand, it contains other latent (internal) stochastic variables $\boldsymbol{\eta}$. In our case, the simulator is the combination of {\tt 21cmFast} and {\tt 21cmSense}, giving rise to multiple realizations of the noisy 21cm power spectra, as detailed in Sec.~\ref{sec:Mock}.  The physical parameters $\boldsymbol{\theta}$ are given in Tab.~\ref{tab:params} while the $\boldsymbol{\eta}$ latent variables are given by the randomly sampled initial conditions of the density and velocity fields. Clearly, the simulator output depends on both $\boldsymbol{\theta}$ and $\boldsymbol{\eta}$,  and typically one is only interested in making inferences on the former by estimating the posterior $p(\boldsymbol{\theta}|{\cal D})$ via the Bayes theorem.

Had we had access to the true likelihood $p({\cal D}|\boldsymbol{\theta})$, we could confidently use MCMC, even if slower, but providing unbiased inference. However, this likelihood is typically intractable, since it is the result of marginalizing over the latent variables: $p({\cal D}|\boldsymbol{\theta}) = \int d\boldsymbol{\eta} ~p({\cal D}|\boldsymbol{\eta},\boldsymbol{\theta}) p(\boldsymbol{\eta}|\boldsymbol{\theta})$. While it is common in the literature to approximate this likelihood somehow (for example, as a Gaussian), the resulting posteriors can show noticeable deviations with respect to likelihood-free approaches \cite{Prelogovic:2023uww}. That being said, for the 21cm power spectrum data the assumption of a Gaussian likelihood is reasonable, at least for the smallest physical scales, while for the highest scales, the Gaussian approximation is expected to break down\footnote{This is all depending on the extent at which the conditions for the Central Limit Theorem applies since the power spectrum is a sum of a large (small) number of contributions at small (large) physical scales.}. Still, the use of SBI is motivated here, since it does not rely on a likelihood approximation, leave apart its computational advantages. 
Note that while in SBI there is no need to evaluate the likelihood, the simulated power spectra coming from {\tt 21cmFAST} that we provide to the algorithm are formally samples from the true (unknown) likelihood, so we do use information of the latter.

In this work, we use a flavour of SBI called Neural-Ratio Estimation (NRE), through the Python library {\tt swyft} \cite{Miller:2022shs}. It aims at estimating the posterior probability $p(\bstheta|{\cal D}_{\rm obs})$, given an observed dataset ${\cal D}_{\rm obs}$, by approximating the likelihood-to-evidence ratio $r({\cal D}_{\rm obs},\bstheta)\equiv p({\cal D}_{\rm obs}|\bstheta)/p({\cal D}_{\rm obs})$ with a neural network model (here $p({\cal D}_{\rm obs})$ is the evidence), and subsequent use of the Bayes theorem to get the posterior via multiplication by the prior. The neural network model is used as a binary classifier, trained to discriminate matching sample pairs $\{{\cal D}_i,\bstheta_i\}$ obtained through the simulation process (i.e., ${\cal D}_i$ is the simulated dataset corresponding to the input value $\bstheta_i$), from those unmatching pairs $\{{\cal D}_i,\bstheta'_i\}$, where $\bstheta'_i$ is unrelated to ${\cal D}_i$ according to the simulator. This makes sense based on the well-known fact that probabilistic binary classifiers (as neural networks) are formally an approximation of the likelihood-ratio between the two concerning hypotheses \cite{2021arXiv211006581H}.
\nnn
The pipeline of our SBI procedure is the following:
\begin{enumerate}[leftmargin=*, labelindent=0pt]
    \item We generate $\bar n=20000$ samples of $\boldsymbol{\theta}$ from $p(\bstheta|{\cal D}_{\rm UV})$ via the MCMC analysis performed in Sec.~\ref{sec:MCMC} consistent with UV luminosity data ${\cal D}_{\rm UV}$. These constitute samples from the prior as far as SBI is concerned and are expected to represent an exhaustive enough coverage of the whole parameter space, for the sake of good reconstruction capabilities (see Sec. \ref{sec:results}). 
\item For each of these values of $\bstheta$, we use the simulator to generate the corresponding (noiseless) simulated dataset ${\cal \tilde D}$. We represent the latter as a 1-dimensional array of power spectrum as a function of $k$, concatenating sub-arrays corresponding to different redshifts, see \cite{Saxena:2023tue} using the same procedure. 
The training data contains $ n= 18 000$ noiseless simulations, while we use the remaining 2000 simulations as validation ($n_{\rm val} = 1 900$) plus test ($n_{\rm test}=100$) data.
We corrupt the training datasets ${\cal \tilde D}$ with ten realisations of thermal noise, thus resulting in a total of $10\times  n=180000$ noisy datasets ${\cal D}_i$, $i=1,..,10\times  n$.
We thus build tuples $\{{\cal D}_i,\bstheta_i\}$ of ``good samples'' (labelled {\tt class 1}).

\item We obtain the same amount of ``bad samples'' (labelled {\tt class 0}) by building tuples $\{{\cal D}_i,\bstheta'_i\}$, where the previously obtained datasets are now paired with random values of the parameters instead, different from that $ \bstheta_i$ used to generate them. 

\item These two sets of tuples will be the input of a classification model, subsequently trained to distinguish {\tt class 1} samples from {\tt class 0} samples. The training consists of minimizing a standard binary ``cross-entropy'' cost function, with the {\tt Adam} stochastic gradient descent algorithm \cite{AdamSGD}. The classifier is given by a feed-forward neural network with three hidden layers, each containing 256 units. The network is trained with a batch size of 64, and we decay the initial learning rate of $10^{-3}$ by 0.95 after every epoch.

\item Once trained, the classifier is used to build the function $r({\cal D}_{\rm obs},\bstheta)$, which together with the prior $p(\bstheta|{\cal D}_{\rm UV})$ can be evaluated in order to obtain the desired posterior $p(\bstheta|{\cal D}_{\rm obs})$. Consistency checks of the resulting posteriors are presented in the form of coverage plots, see Appendix \ref{app:coverage} for the results. The interested reader can find more details in \cite{TMNRE}.

\end{enumerate}

\section{Results}
\label{sec:results}

In this section, we present the results of our study. In Sec.~\ref{sec:recovery}, we show the generic reconstruction capabilities of our SBI algorithm in terms of marginal (single parameter) posteriors. Given the high computational cost of a single {\tt 21cmFast} simulation, we include as well a comparison of the reconstruction capabilities when the number of simulations, $n'$, used for training decreases, i.e. when $n'< n$. In Sec.~\ref{sec:2Dposteriors}, we discuss the reconstruction of the DM mass in the presence of degeneracies or correlations with the other astrophysics parameters analyzing the corresponding 2D posterior distributions. We close the discussion presenting the $\mwdm$ bound resulting from our analysis as a function of $\Mt$ in Sec. \ref{sec:bounds}.

\subsection{Parameter reconstruction}
\label{sec:recovery}

\begin{figure}[t!]
\centering
\includegraphics[width=.49\linewidth]{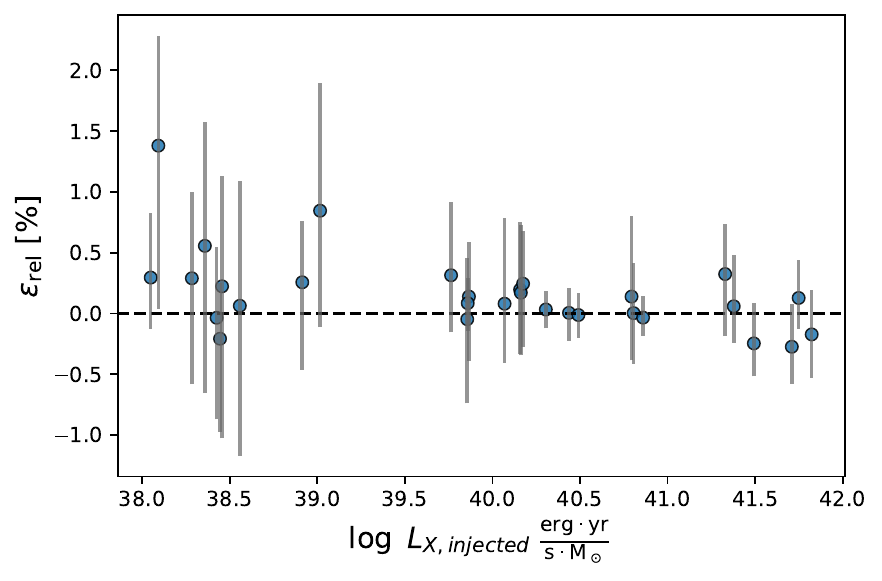}
\includegraphics[width=.49\linewidth]{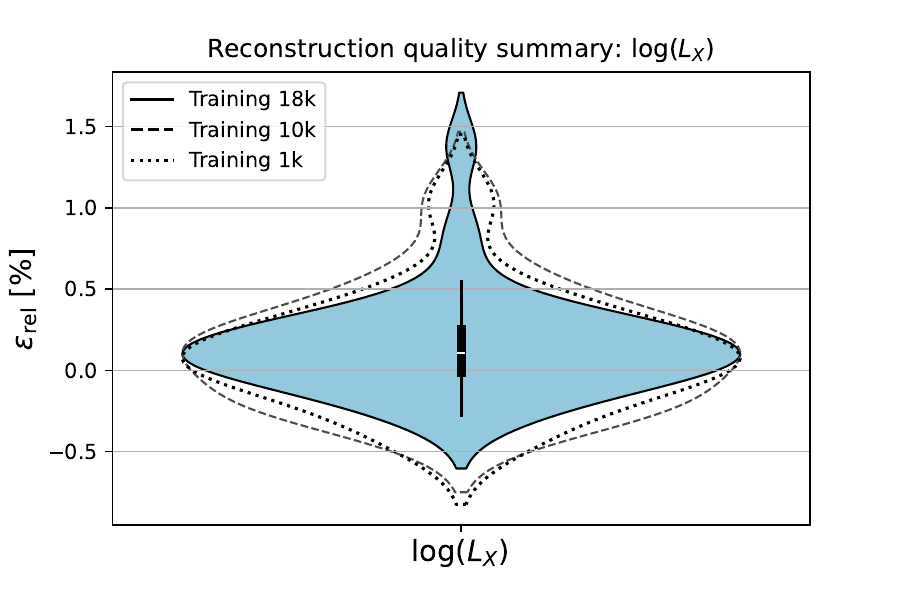}
\includegraphics[width=.49\linewidth]{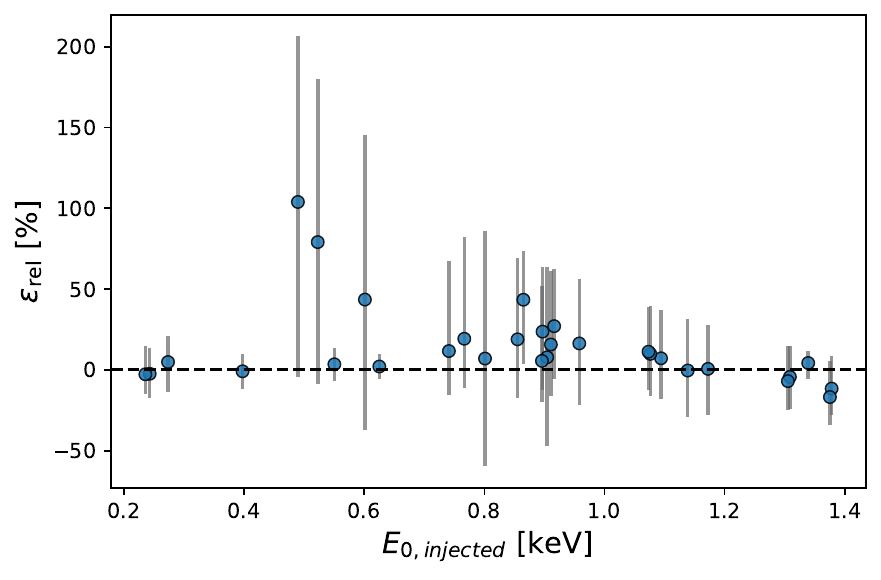}
\includegraphics[width=.49\linewidth]{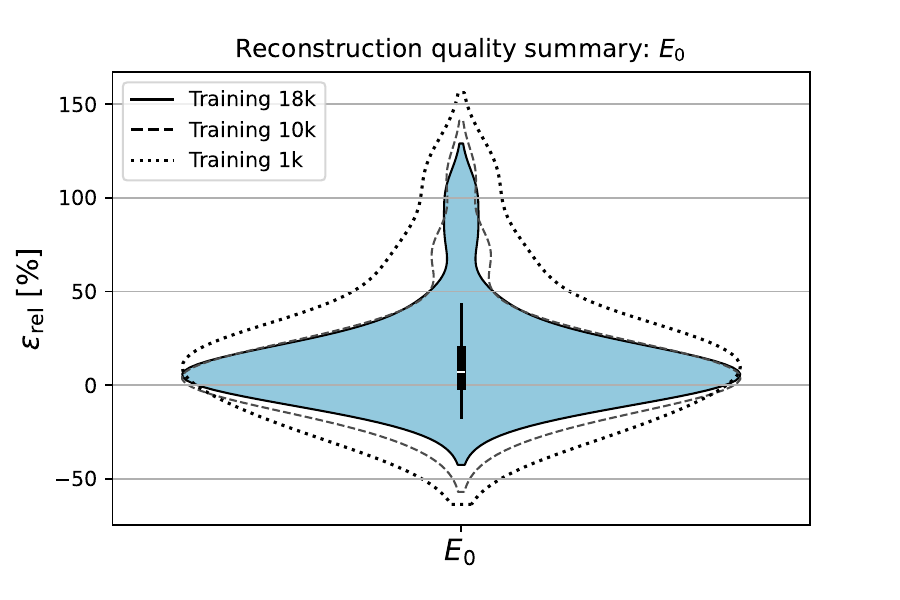}
\caption{Reconstruction quality in terms of relative error $\epsrel$ for the X-ray parameters used in this work. Left panel: relative error in the reconstruction as a function of the injected value of the parameter for $N_{\rm sim}=30$ random simulations. The error bars represent the 95\% CI of the posterior distribution in each case. Right panels: corresponding distribution of the relative error for a training setup with $n=18$k simulations (blue violins). The transparent violins are analogous to the blue one, but using instead $n'=10$k simulations (dashed) and $n'=1$k simulations (dotted). See text for details.  }
\label{fig:key_violin_Xray}
\end{figure}

\begin{figure}[t]
\centering
\includegraphics[width=.49\linewidth]{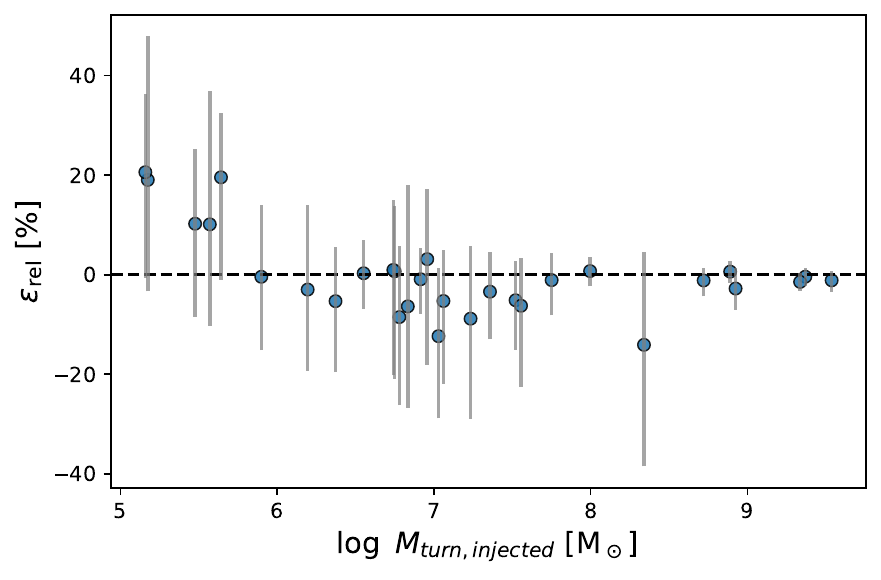}
\includegraphics[width=.49\linewidth]{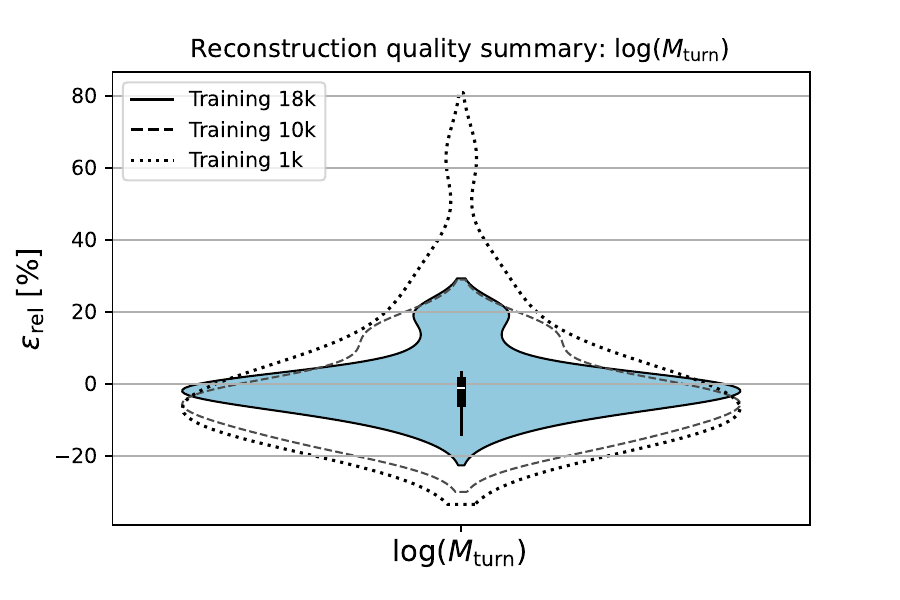}
\includegraphics[width=.49\linewidth]{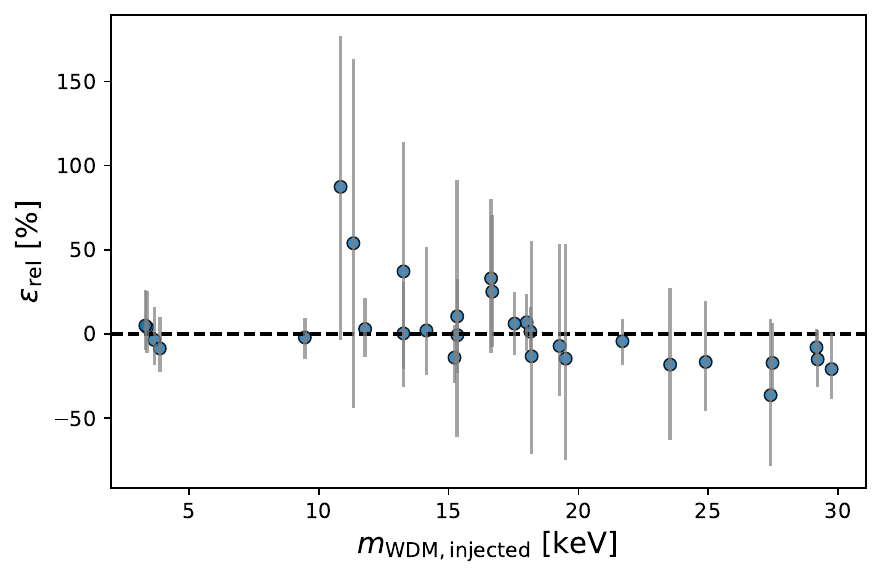}
\includegraphics[width=.49\linewidth]{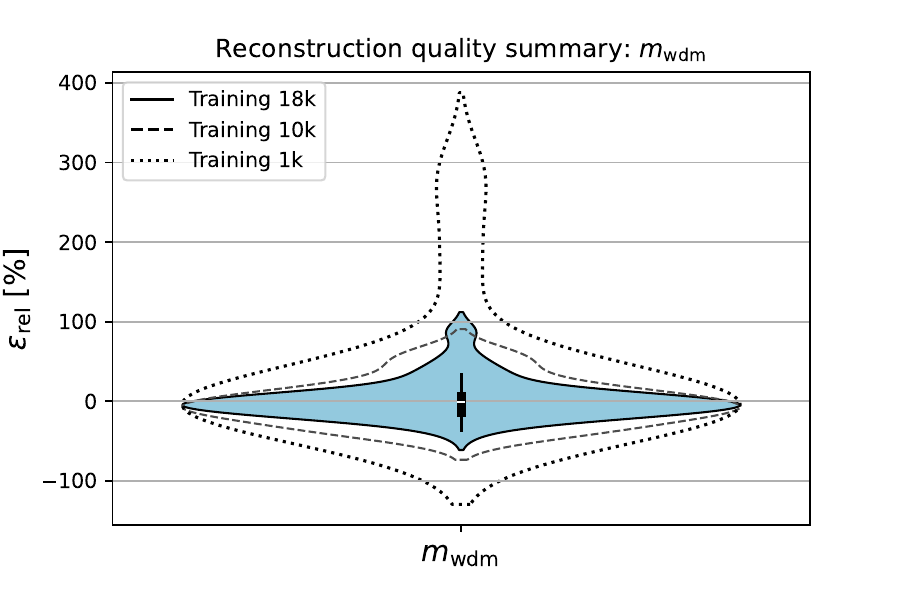}
\caption{ Analogous to Fig.~\ref{fig:key_violin_Xray} but for the the $\mwdm$ and $M_{\rm turn}$ parameters. The 30 points in each plot correspond to the same 30 random simulations used in Fig.~\ref{fig:key_violin_StarF}.}
\label{fig:key_violin_MM}
\end{figure}

\begin{figure}[t!]
\centering
\includegraphics[width=.49\linewidth]{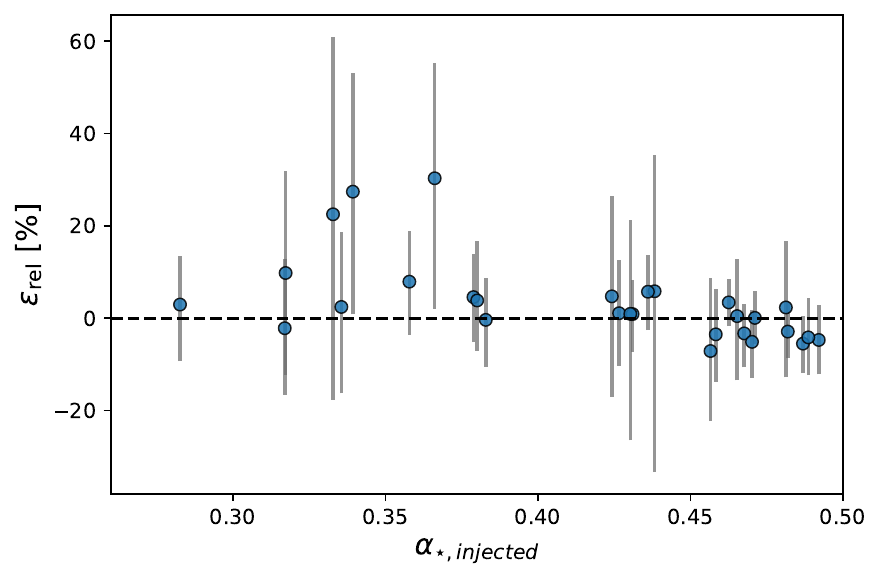}
\includegraphics[width=.49\linewidth]{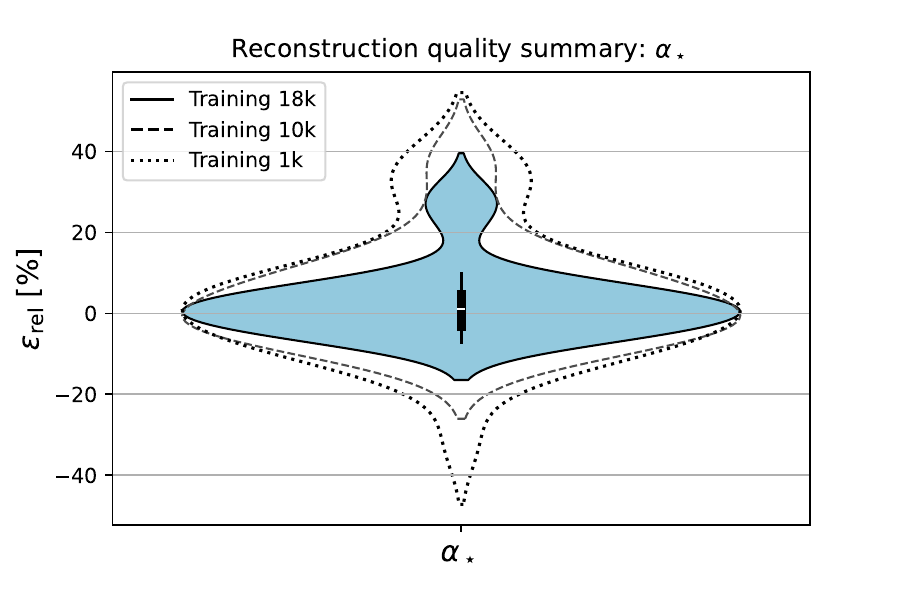}
\includegraphics[width=.49\linewidth]{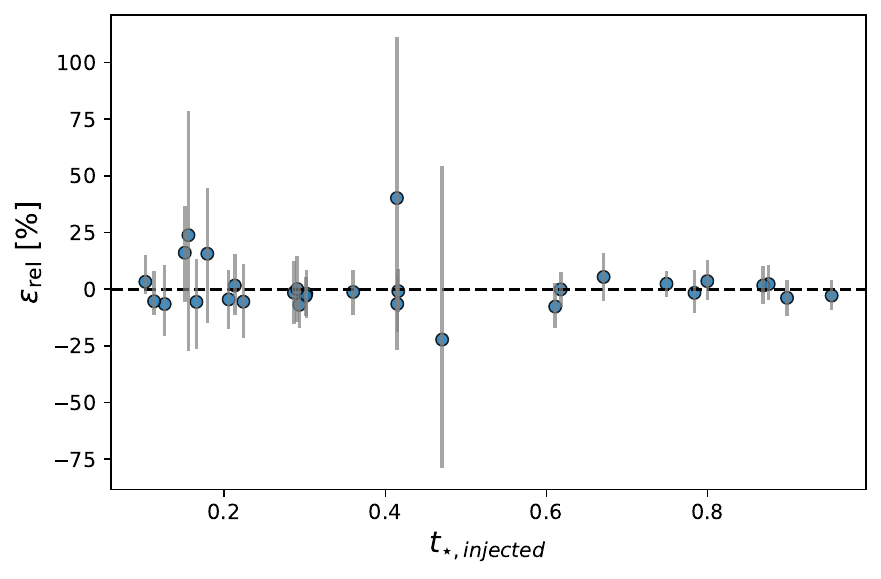}
\includegraphics[width=.49\linewidth]{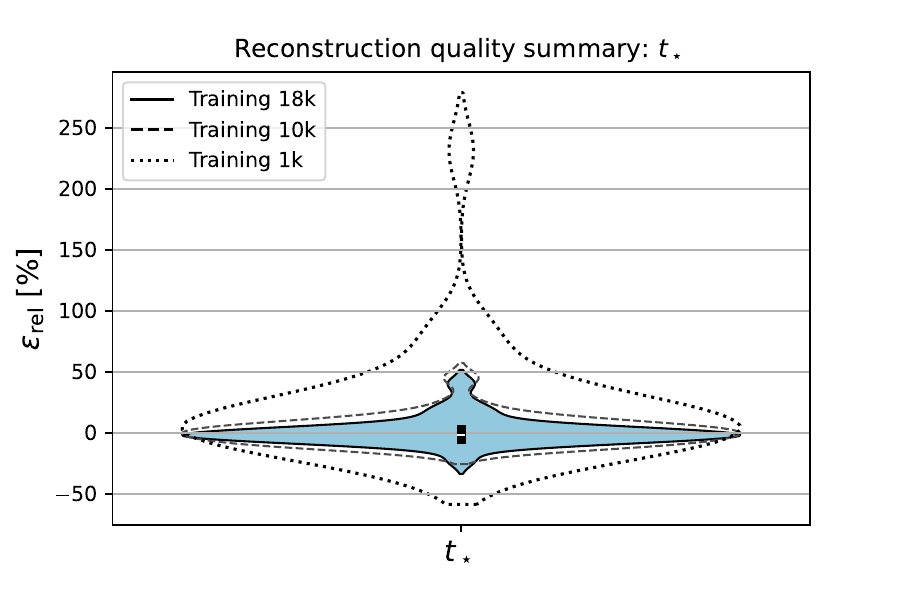}
\includegraphics[width=.49\linewidth]{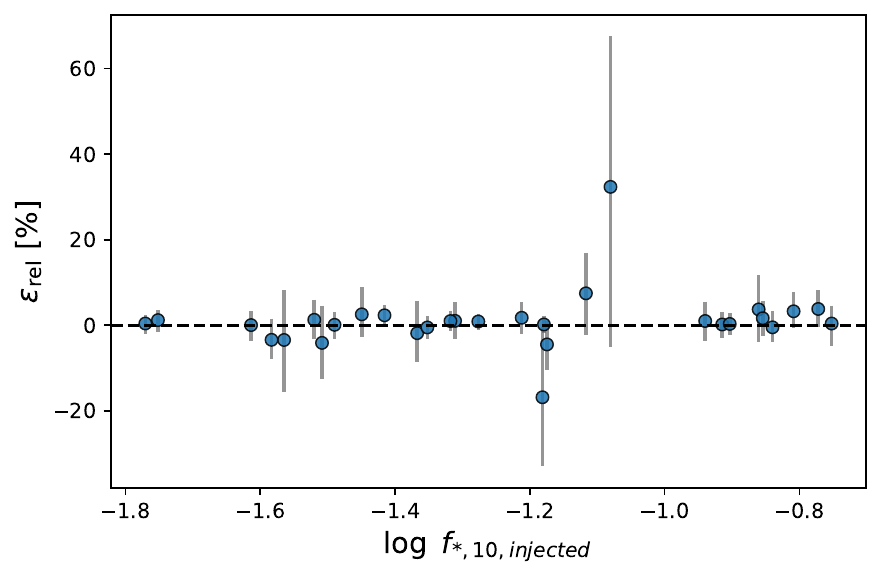}
\includegraphics[width=.49\linewidth]{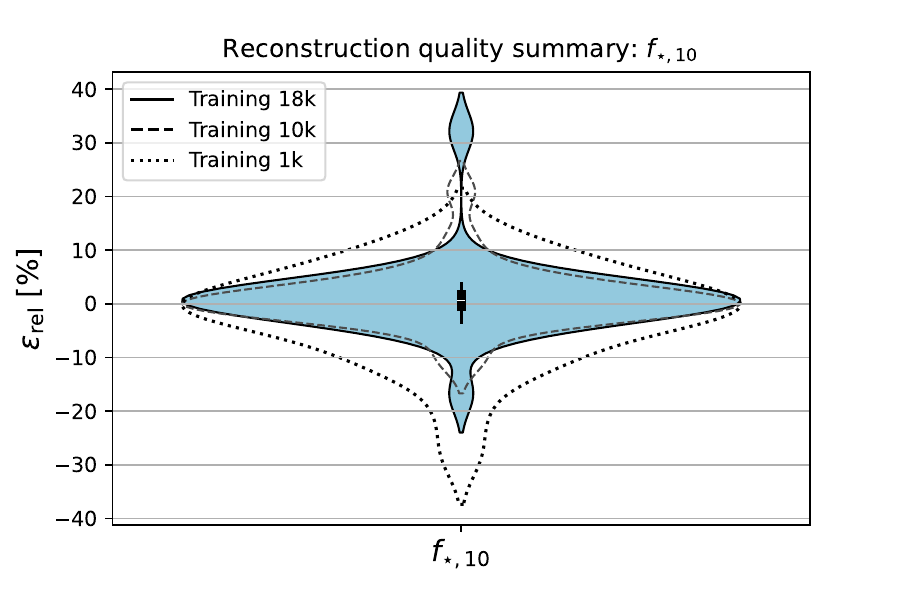}
\caption{ Analogous to Fig.~\ref{fig:key_violin_Xray} but for the SFR  parameters considered in this work. The 30 points in each plot correspond to the same 30 random simulations used in Fig.~\ref{fig:key_violin_Xray}.
}
\label{fig:key_violin_StarF}
\end{figure}

\begin{figure}[t!]
\centering
\includegraphics[width=0.99\linewidth]{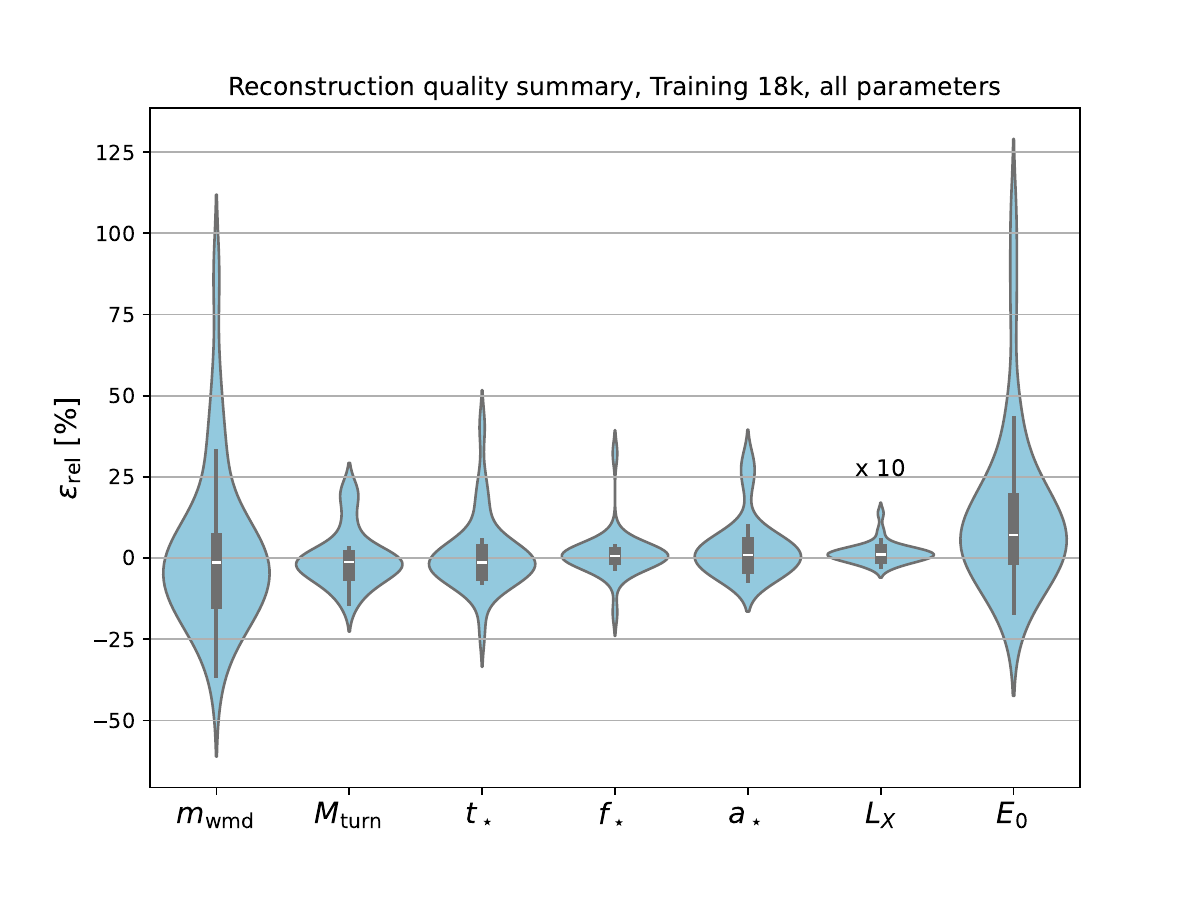}

\caption{Summary of the violin plots representing the distribution of the relative error $\epsrel$ when considering  $n= 18000$ simulations in the reconstruction of the parameters reported at the bottom of the figure. Note that the violin for $L_X$ has been rescaled by a factor 0.1 for visibility purposes. A zoom-in version of those violins and their degradation reducing the number of simulations were shown in the previous Figs.~\ref{fig:key_violin_Xray},~\ref{fig:key_violin_MM} and~\ref{fig:key_violin_StarF}.}

\label{fig:all_violins}
\end{figure}

We first discuss the performance of our procedure in reconstructing the different parameters as follows:
\begin{enumerate}[leftmargin=*, labelindent=0pt]
    \item  For each parameter, we compute the marginal posteriors corresponding to each of $N_{\rm sim}=30$ simulations. The latter has been chosen randomly from the set of $n_{\rm test}$ simulations that were kept apart from the ones used to train our algorithms. The posterior of every parameter $\theta$ gives rise to an expected value, $\langle \theta\rangle$, which we compare to the true value $\theta_{\rm true}$ of the corresponding simulation to construct the relative error, $\varepsilon_{\rm rel} = 100(\theta - \theta_{\rm true})/\theta_{\rm true}$, 
as a function of $\theta_{\rm true}$. The results are shown in the left panels of Figs.~\ref{fig:key_violin_Xray}, \ref{fig:key_violin_MM}, and \ref{fig:key_violin_StarF} where the error bars correspond to the 95\% CL of the posteriors.
    \item From the above procedure, we have $N_{\rm sim}$ values of $\varepsilon_{\rm rel}$ for each parameter $\theta$, whose distributions are shown in violins plots in the right panels of the same figures. Blue violins are shown together with their corresponding ``candle'' in the centre. The latter represents the mean (central white marker), the Interquartile range (IQR, spanning from the 25\% to the 75\% quartile) at the extremes of the thick bar, as well as the points extending up to 1.5 times the IQR (thin inner vertical segment). All those violins are reported in Fig.~\ref{fig:all_violins} to ease their one-to-one comparison.
    
    Given the high computational cost of running {\tt 21cmFAST} in our machines ($\sim 20'$ per simulation to obtain the power spectrum), it is natural to ask how the reconstruction capabilities degrade as we decrease the number of simulations used for training. While so far the results have used all the $n=18$k available simulations (blue violins), we repeat the exercise with $n'= 10$k simulations (dashed transparent violins) and $n'= 1$k simulations (dotted transparent violins).

\end{enumerate}

\noindent
Figure~\ref{fig:all_violins} highlights different reconstruction capabilities across the parameters. The best reconstruction is achieved for the parameter $L_X$, see the upper panels of Fig.~\ref{fig:key_violin_Xray} for more details. We see that the  $\epsrel$ is indeed below the per cent level in most cases. Note that actually, the first results of HERA \cite{HERA:2021noe} were already able to constrain this parameter.  In the case of $E_0$, the reconstruction quality worsens for the true values of $E_0\gtrsim 0.6$ keV, as visible in the bottom panels of Fig.~\ref{fig:key_violin_Xray}. Similar behaviour has been observed in \cite{Schosser:2024aic} for a different parametrization of the 21cm signal, implementation of the WDM imprint and a different range of the corresponding parameters, see the discussion in ~Sec.~\ref{sec:introduction}. Here in particular we consider a wider range of $\mwdm$ and $\Mt$ values.
Note that larger values of the energy threshold for X-rays escaping galaxies, $E_0$,   make the X-ray photon spectrum heating the IGM harder. Harder X-rays have longer mean free path, $\lambda_X\sim E_X^{2.6}$ where $E_X$ denotes the X-ray energy~\cite{Mesinger:2012ys}.  This implies that for larger $E_0$ the heating is happening in a more homogeneous manner leading to less contrast in the 21cm signal at EoH when other parameters are fixed. Consequently, we have a suppressed power spectrum at EoH and thus a more difficult reconstruction of $E_0$ for these higher values.

On the other hand, focusing now on the $M_{\rm turn}$ parameter in the upper panels of Fig.~\ref{fig:key_violin_MM} we observe that the reconstruction quality is very high for the largest true values of this parameters while becoming more challenging for $M_{\rm turn}\lesssim 10^{7.5}M_\odot$. This is expected, since  $M_{\rm turn}>10^{7.5}M_\odot$ is expected to drive the overall shift of the power spectrum to later redshifts as our largest considered WDM effect induces an $\Mcut=2\times 10^7 M_\odot$ for $\mwdm=3$ keV. Lower $M_{\rm turn}$ values have an imprint that is more difficult to disentangle from other parameters e.g. the free-streaming effect of  WDM  could prevent a good reconstruction of the former. We have checked that this is the case for the two left-most points in the bottom-left panel of Fig.~\ref{fig:key_violin_MM},  corresponding to a very low $\mwdm$. Overall, the reconstruction error of $M_{\rm turn}$ is always better than $\sim 20\%$ for the set of simulations that we have used. On the other hand, the reconstruction of the DM mass is much better at the lowest $\mwdm$, see the  bottom panels of Fig.~\ref{fig:key_violin_MM}.  This is for reasons analogous to the better reconstruction of large $M_{\rm turn}$, i.e. the impact of a sufficiently low $\mwdm$ on the power spectrum can be more easily distinguished from the effect of the other parameters. Note that, at the highest DM masses considered in this example, the $\epsrel$  is also reasonably good. We have checked that this is due to relatively low true values of $M_{\rm turn}$ for our particular choice of simulations.

On the other hand, we see in Fig.~\ref{fig:key_violin_StarF} that the SFR parameters are reconstructed at the ${\cal O}(10-50\%)$ level. Yet,   their $\varepsilon_{\rm rel}$ distributions clearly peak around zero, for the other parameters. We observe a better reconstruction in general for the largest true values of the SFR parameters. This is expected as, for example, in the case of $\alpha_\star$ (upper panels), sufficiently large values of this parameter will imply larger star-forming halos, which in turn becomes the dominating cause for the delay in the appearance of the power spectrum features, compared to the effect of the other parameters. Similar reasoning can be made for $t_\star$ (central panels), controlling the SFR time scale. Larger $t_\star$ values imply a suppression of the SFR, which causes a delay in the power spectrum features.

Concerning the degradation of the reconstruction when training with a smaller number of simulations, we can observe in the right panels of Figs.~\ref{fig:key_violin_Xray},~\ref{fig:key_violin_MM} and~\ref{fig:key_violin_StarF} that the lowest-budget training of $n'= 1$k noiseless simulations is not enough to converge on the reconstruction, thus producing a much wider distribution of $\varepsilon_{\rm rel}$ for all the parameters. In contrast, the intermediate-budget training,  using $n'= 10$k simulations, produces in general a reconstruction with similar quality to the one with the highest-budget training ($n=18$k). As a result,  we can conclude that using the {\tt 21cmFast} simulator and the SBI technique that we have adopted for our setup, the minimum number of simulations to produce good quality results is 10k.

\subsection{Degeneracies in the posterior distributions }
\label{sec:2Dposteriors}

Having presented the generic reconstruction capabilities of our algorithm, here we address the degeneracies between the DM mass and the other astrophysics parameters discussed in Sec.~\ref{sec:WDM21}. We do so by considering the 2D posterior distributions for one illustrative fiducial scenario and identifying the shape of the different contours.

\begin{figure}[t!]
\centering
\includegraphics[width=0.99\linewidth]{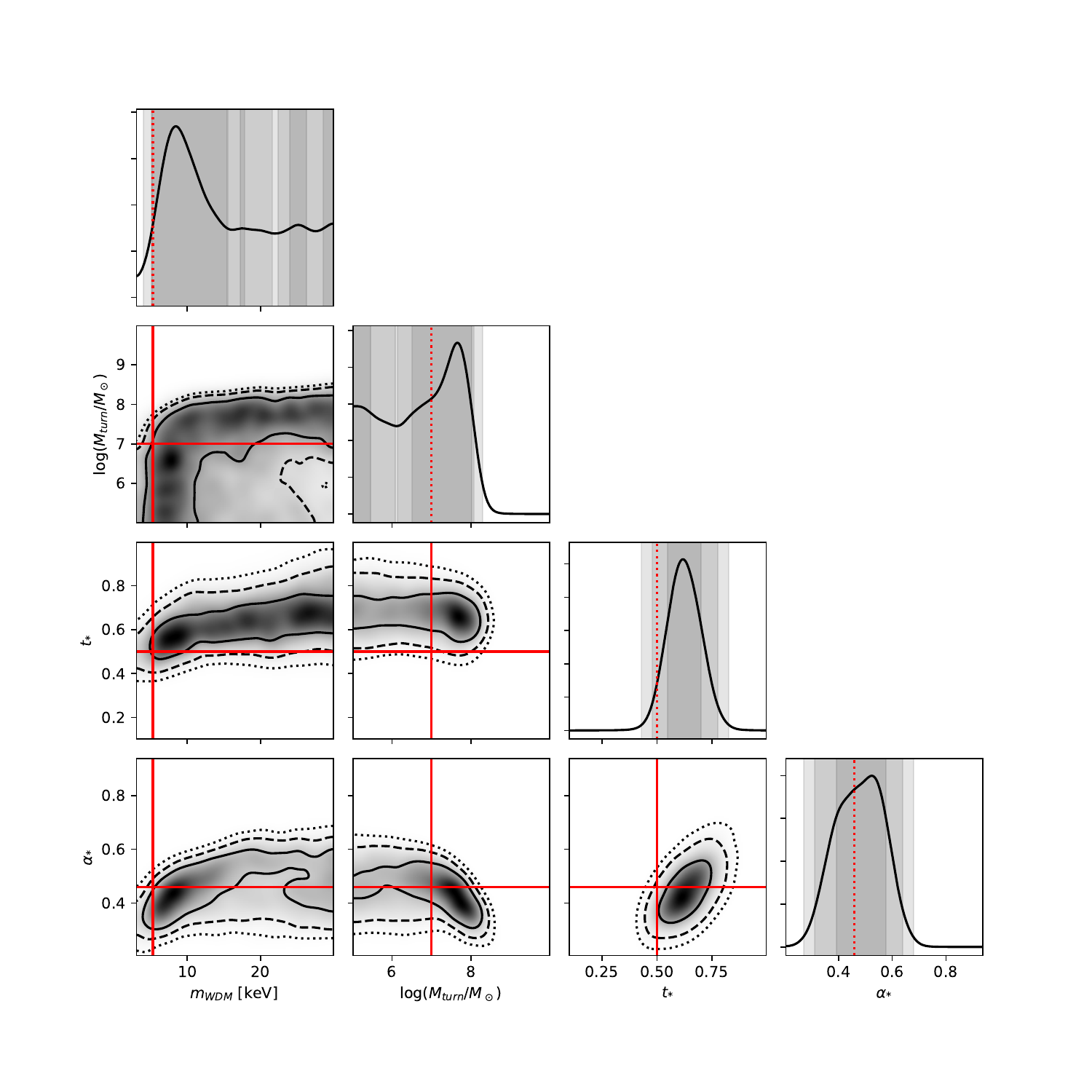}
\caption{Reduced set of recovered 1D and 2D marginalized posteriors on our astrophysics and WDM  parameters assuming a mock 1000h observation of the 21-cm power spectrum with HERA for $\mwdm=5.3$~keV and $M_{\rm turn}=10^7 M_\odot$. The other parameters have been fixed to the same fiducial values as in Fig.~\ref{fig:PSnonoise}. The entire triangle plot, including all the free parameters considered in our SBI analysis, is shown in Figure~\ref{fig:triangle-full-9}. }
\label{fig:triangle-reduced}
\end{figure}

In Fig.~\ref{fig:triangle-reduced}, we focus on the correlation between $\mwdm$ and $M_{\rm turn}, \alpha_\star,t_\star$ to ease the reading of the plot. Indeed, it appears that the HERA sensitivity is such that no strong correlation appears between $\mwdm$ and $\fst, L_X, E_0$, see  the full triangle plot in Fig.~\ref{fig:triangle-full-9} in the appendix.
For illustrative purposes, we have fixed $\mwdm=5.3$ keV and $\Mt=10^7 \,M_\odot$ for the fiducial scenario and we have fixed all the other parameters to the same values as in Figs.~\ref{fig:PSnonoise} and~\ref{fig:PSnoise}. Since our choice of $\Mt$ is such that  $\Mt\sim  \Mcut(5.3 {\rm keV})$, we expect that degeneracies shall appear between the $\mwdm$ and $\Mt$. Furthermore, from the discussion in the previous section, we expect
a relatively good recovery of low dark matter mass with HERA sensitivity. All parameters shall thus be relatively well reconstructed for our choice of fiducial scenario. 

These expected trends can indeed be observed in the posterior distributions of Fig.~\ref{fig:triangle-reduced}. The top panels correspond to the marginalised 1D posteriors whereas the other panels show the 2D posteriors. The continuous, dashed and dotted contours in the 2D panels correspond to the 68\%, 95\% and 99\% confidence levels (CL) contours. The same percentiles are represented as darker, medium and lighter grey regions in the marginalised 1D posteriors. The vertical and horizontal red lines indicate our fiducial parameter values. As visible here and in the full triangle plot, the true values of the parameters are always lying within the 95\% CL intervals confirming the relatively good reconstruction of the parameters.   Furthermore,
a positive correlation between $\mwdm$ and $M_{\rm turn}$ is visible, showing the expected behaviour, see Sec.~\ref{sec:WDM21}. Indeed, to compensate for an increase in $M_{\rm turn}$, which shifts the power spectrum features to later times, one could consider colder dark matter by increasing the WDM  mass (and vice versa). For sufficiently low WDM mass, low $M_{\rm turn}$ cannot be discarded, since the power spectrum would be insensitive to the latter; resulting in a high tail at low $M_{\rm turn}$ values as seen in the corresponding 1D distribution. The same logic applies to the WDM mass: for sufficiently high $M_{\rm turn}$, high values of $\mwdm$ will have no impact on the power spectrum, thus resulting in a high tail at high $\mwdm$, shown in its 1D posterior.


An analogous reasoning can be made when looking at the degeneracy between $\mwdm$ and the other two parameters shown in Fig.~\ref{fig:triangle-reduced}.  To compensate for an increase in $\alpha_\star$, leading to a larger fraction of stars into heavier haloes, we should increase $\mwdm$ so as not to suppress low mass haloes population. On the other hand, when $t_\star$ increases the SFR decreases, which is again compensated by a colder DM, i.e. larger $\mwdm$, thus featuring a similar positive correlation among these two parameters.

\subsection{Impact of $M_{\rm turn}$ on the  $\mwdm$  lower bound}
\label{sec:bounds}

We now turn our attention to the potential for 21cm Cosmology, in the particular case of the HERA telescope, to surpass existing probes of the warmness of dark matter. More concretely, we aim to put lower bounds on the WDM mass as a function of the injected threshold mass $M_{\rm turn}$, while keeping fixed the rest of the astrophysical parameters. The injected $\mwdm$ mass has been fixed to $28$ keV, as a representative of a CDM scenario, while not being right at the extreme of our prior range to avoid border effects. Following the standard procedure in Bayesian statistics, for every value of the injected $M_{\rm turn}$ we obtain the 95\% CL bounds, $\mwdm^{\rm min,95}$, by solving for the latter in the equation
\[
0.95 = \int_{\mwdm^{\rm min,95}}^\infty d\mwdm~p(\mwdm|{\cal D})~,
\]
where $p(\mwdm|{\cal D})$ is the $\mwdm$ posterior distribution given the data ${\cal D}$. Such data consists of 5 simulations ${\cal D}=\{{\cal D}_1, {\cal D}_2,...,{\cal D}_5\}$, each with different initial conditions but generated from identical fiducial parameter values, and including thermal noise. The latter  was sampled from HERA sensitivities using {\tt 21cmSense} as in the training procedure. We make inference on each of these simulated datasets using our NRE algorithm, previously trained on the maximum budget of $n=18$k simulations, see  Sec.~\ref{sec:recovery}. The combined posterior can thus be  simply estimated as:\footnote{We assume that all the ${\cal D}_{1,...,5}$ are independent.}
\begin{equation}
p(\mwdm|{\cal D}) = p(\mwdm)~\prod_{i=1}^5
r_i(\mwdm;{\cal D}_i)~,
\label{eq:combined_posterior}
\end{equation}
where $p(\mwdm)$ is the DM mass prior, and $r_i(\mwdm;{\cal D}_i)$ is the output of the algorithm for the dataset ${\cal D}_i$, noting that this is nothing but the estimation of the likelihood-to-evidence ratio.

\renewcommand{\arraystretch}{1.15}
\begin{table}[t]
	\centering
	\begin{tabular}{ | c | c  c  c  c c c|  }
		\hline
        benchmark & $\log(\fst)$&$\as$& $\ts$&  $L_X$& $E_0$& $\mwdm$\\ 
		& & & & [$\LXu$]& [keV]& [keV]\\
		\hline \hline
           1&-0.9 & 0.46 & 0.5 & $10^{40}$ & 0.5 & 28 \\
              2&-0.9 & 0.46 & 0.5 & $10^{41} $& 1.1 & 28\\
		\hline
	\end{tabular}
	\caption{ Benchmark models submitted to the trained NRE to extract the 95\%CL bounds on $\mwdm$ as a function of $\Mt$ in Figs.~\ref{fig:sensitivitySBI} and~\ref{fig:sensitivitySBIb2}. }
	\label{tab:bench}
\end{table}

\begin{figure}[t!]
\centering
\includegraphics[width=0.85\linewidth]{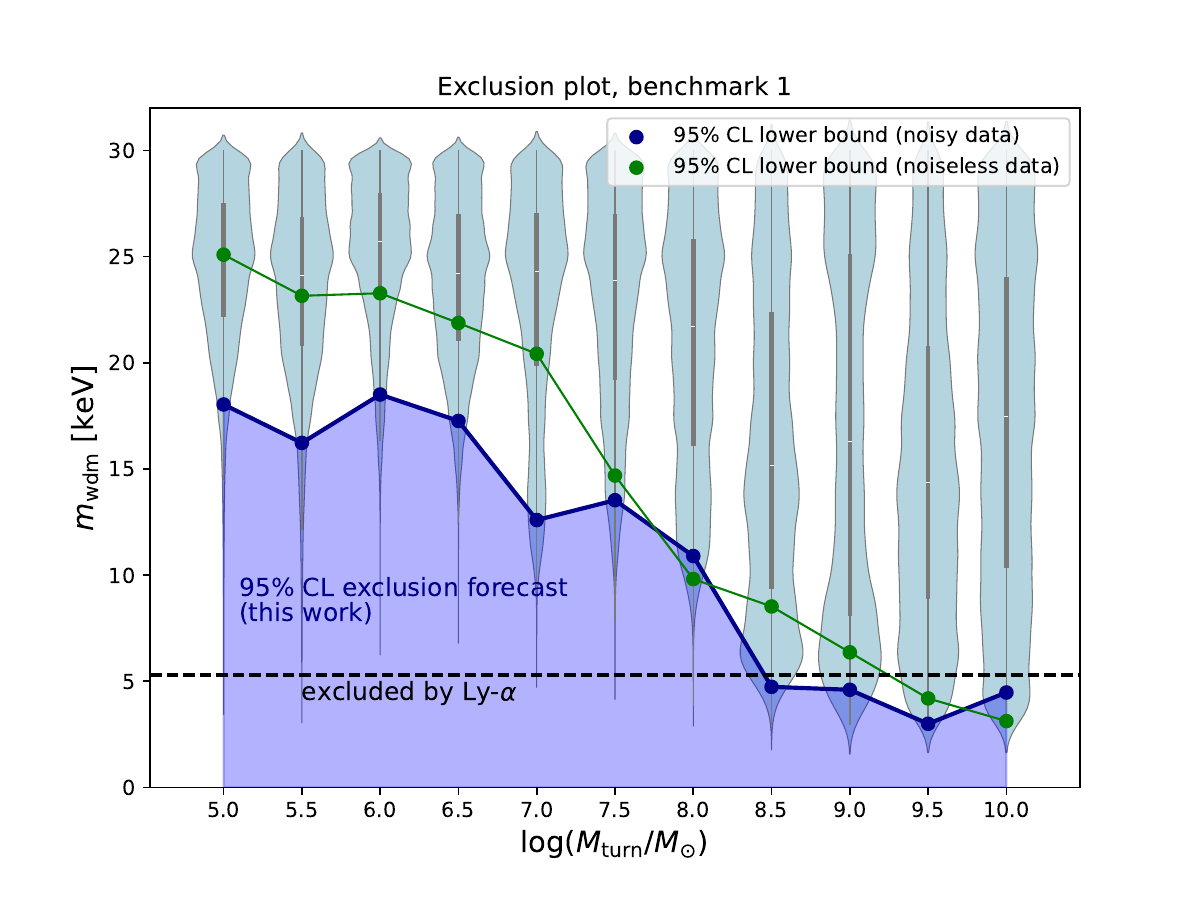}
\caption{ Forecast of the HERA sensitivity  to $\mwdm$ as a function of the injected $M_{\rm turn}$ value. For every $M_{\rm turn}$ we show in blue violins the posterior distribution of $\mwdm$ from 5 noisy simulations of benchmark 1,  see Tab.~\ref{tab:bench}. The corresponding 95\% CL lower bound on $\mwdm$ is shown with the dark blue bullets, while the analogous lower bound using noiseless data is shown with green bullets. The lower bound from Ly-$\alpha$ forest data is included as a dashed line for comparison. }
\label{fig:sensitivitySBI}
\end{figure}

\begin{figure}[t!]
\centering
\includegraphics[width=0.85\linewidth]{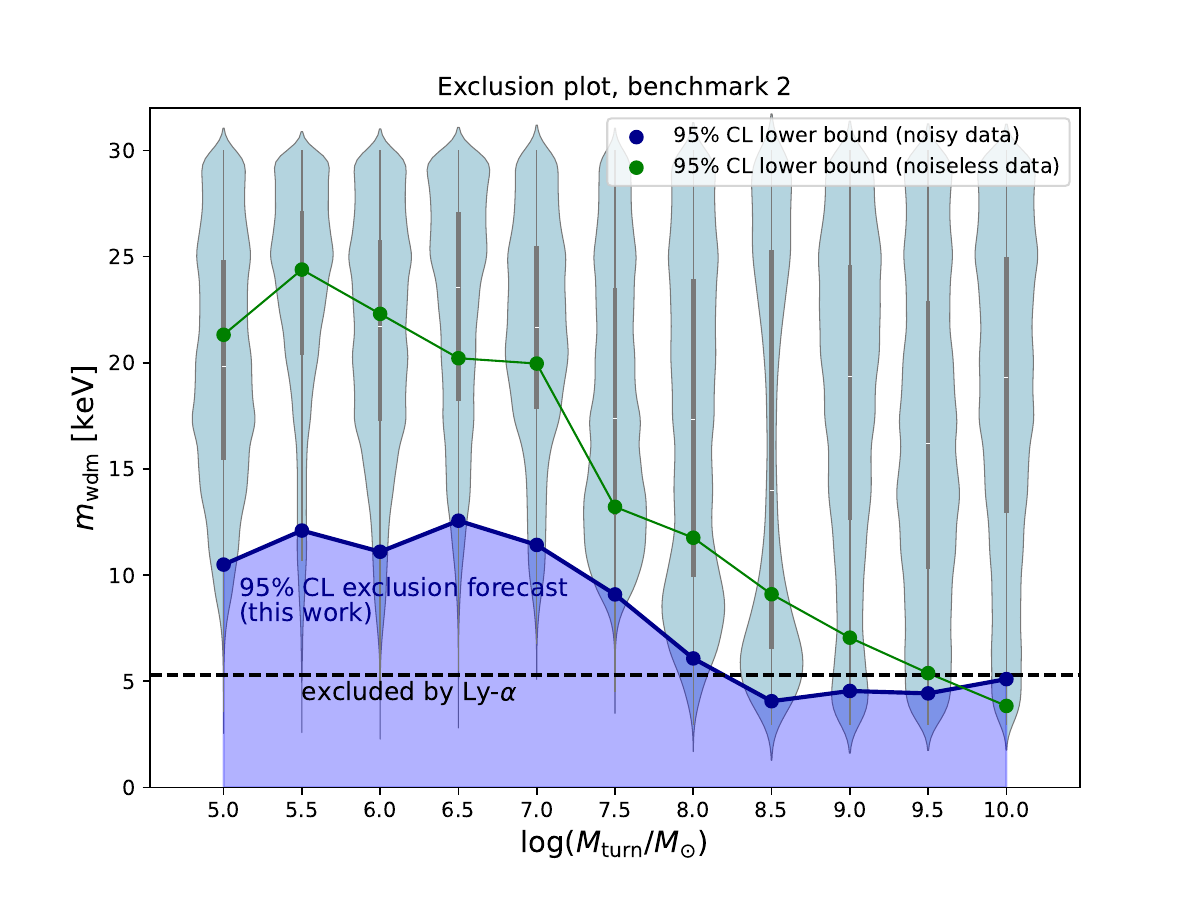}
\caption{ Same as Fig.~\ref{fig:sensitivitySBI} but for benchmark 2 displaying higher values of the X-ray parameters, see Tab.~\ref{tab:bench}. }
\label{fig:sensitivitySBIb2}
\end{figure}

We compute the bound for a regular grid of $M_{\rm turn}$ masses in the range $[10^5 M_\odot, 10^{10} M_\odot]$. The results are shown in Figs. \ref{fig:sensitivitySBI} and~\ref{fig:sensitivitySBIb2} for two benchmarks, specified in Tab.~\ref{tab:bench}.
We consider those two benchmarks to illustrate the dependence on the bound both on $\Mt$ and on the X-ray parameters. Benchmark 2, in particular, exhibits higher $E_0$ and $L_X$ values, which shall result in a more homogeneous heating of the IGM. Consequently, a more suppressed noiseless power spectrum is expected within the redshift range probed by HERA. With all other parameters fixed, this makes benchmark 2 potentially more challenging to test. In both figures, each violin represents the posterior distribution $p(\mwdm|{\cal D})$ for a given injected $M_{\rm turn}$, reported on the x-axis, while the blue dots represent the corresponding 95\% lower bound on $\mwdm$. On the other hand, for both benchmarks we repeat the exercise by using instead {\it noiseless} data, i.e. not adding thermal noise, intuitively leading to stronger WDM bounds with respect to the analogous inference on noisy data. The results, represented with the green dots, are just for academic purposes. They indicate how much the thermal noise degrades the bounds.

As expected from the discussion in Sec.~\ref{sec:TKdndM}, larger threshold masses would prevent a good reconstruction of the DM mass, resulting in weaker bounds on the latter.  Nonetheless, for both benchmarks, we find that the lower bound on $\mwdm$ can surpass the Lyman-$\alpha$ forest constraint of 5.3 keV (horizontal black dashed line in both figures) if $\Mt$ is below $10^{8} \,M_\odot$. Furthermore, if all CD galaxies have $\Mt = 10^{5} \,M_\odot$, or equivalently, if all CD galaxies are assumed to be in the form of MCGs, thermal DM masses up to $\mwdm \simeq 17$ keV and  $\mwdm \simeq 10$ keV could be excluded with the X-ray parameters of benchmark 1 and 2, respectively.
Even if it might be unrealistic to consider that all CD galaxies could be part of one single population of galaxies, the latter number provides us with a quantitative {\it lower limit} on the  WDM mass that could be probed with HERA for our choices of fiducial scenario. 
As expected, in the noiseless case (green dots), the resulting lower bounds on $\mwdm$ are stronger than in the noisy case (blue dots).

One could also provide an analysis involving multiple galaxy populations. An instance of the latter is available in {\tt 21cmFast} combining MGC and AGC while accounting for feedback effects, see~\cite{Qin2020MNRAS.495..123Q}. Yet, this implies much slower simulation and a much larger set  of parameters to account for the two population galaxy properties or, equivalently, a larger number of simulations to run in order for the training procedure to converge. Furthermore,  this would induce extra degeneracies between those parameters and the WDM mass. Such an analysis is beyond the scope of this paper. Our main objective was to emphasize that the choice of threshold mass for star-forming galaxies, or equivalently,  the choice of galaxy populations plays a major role in the determination of the sensitivity of 21cm Cosmology experiments on the NCDM imprint. Similar conclusions were reached in~\cite{Giri:2022nxq} providing a sensitivity estimate of SKA on a discrete set of source modeling below virial masses of  $10^8 M_\odot$.  In addition, comparing Figs.~\ref{fig:sensitivitySBI} and ~\ref{fig:sensitivitySBIb2}, for the selected X-ray parameters considered, we have a rather mild effect on the maximal threshold mass that would help to improve on existing bounds on $\mwdm$. In contrast, the lower bound $\mwdm$ at the lowest threshold masses considered in this analysis could become less stringent for the higher explored values of the X-ray parameters giving rise to a more suppressed power spectrum at the redshifts of interest.

\section{Conclusion}
\label{sec:conclusion}

The redshifted 21cm signal from CD and EoR will soon be probed to a high signal-to-noise ratio thanks to the development of radio interferometers,  such as HERA and SKA.  These are expected to probe the distribution of IGM fluctuations at those epochs, inaccessible as of today. The 21cm signal will depend on -yet unknown- astrophysics phenomena and cosmic dawn galaxy evolution together with potential distinctive imprints beyond the standard model (BSM) phenomena. In this paper, we go beyond the $ \Lambda$CDM model considering the possibility for dark matter to be non-cold. As is well known, a plethora of BSM scenarios can give rise to an NCDM behaviour, yet, for concreteness we focus on thermal Warm Dark Matter as a benchmark for free-streaming dark matter. 

Thermal WDM is expected to significantly suppress the number density of low halo masses, giving rise to a delay in the 21cm signal features. In this work, we parametrize the WDM imprint on the linear matter power spectrum by making use of a  transfer function parametrized as in  eq.~(\ref{eq:TWDM}) involving a free-streaming scale described in eq.~(\ref{eq:alphWDM}). Furthermore,  for the non-linear evolution of the matter density perturbations, we have updated the WDM effect on the halo mass function in the public code {\tt 21cmFast} by considering a sharp-k cutoff window function instead of the default top-hat one, see Sec.~\ref{sec:WDM21}. Our astrophysics framework is the cosmic dawn galaxy parametrization introduced in~\cite{Park:2018ljd} considering one single galaxy population characterized by a threshold mass for star formation $M_{\rm turn}$, a star formation rate amplitude, $f_{\star,10} $, and tilt $\alpha_{\star}$. The X-ray heating from stars is driven by the soft X-ray luminosity amplitude, $L_X$, and the energy threshold for X-rays to escape galaxies, $E_0$. All these astrophysics parameters shape the 21cm signal and can be expected to display some level of degeneracy with the WDM mass, $\mwdm$, that controls the WDM free streaming length. In this paper in particular, we emphasize the key role played by the $M_{\rm turn}$ parameter in the sensitivity of the 21cm signal to the WDM imprint.

Traditionally, likelihood-based inference methods, and especially MCMC,  have been used in cosmology, in general. Apart from relying on a good likelihood specification (something which is not possible in general), MCMC methods have other features which render them computationally inconvenient. First,  they require to produce joint samples of the full parameter space (even beyond those parameters of direct interest). Second,  a full MCMC from scratch has to be performed for every newly observed dataset. In our particular case of the 21cm power spectrum, the likelihood may be Gaussian to a good approximation. Yet, the computational disadvantages mentioned above still apply. Consequently, here we address these problems by making use of a Simulation-based Inference (SBI) analysis, which does not rely on a likelihood specification, while being on the other hand advantageous from the computational point of view with respect to MCMC methods. 

We generate mock data with {\tt 21cmFast} while varying the initial conditions from a random seed to account for the cosmic variance. Before our Bayesian inference on such data, we constrain the parameters of interest from the UV luminosity data from~\cite{Bouwens2021AJ}, as a result of which we obtain priors to our SBI analysis. Our MCMC results reported, in Fig.~\ref{fig:MCMC}, provide qualitatively similar results to~\cite{Park:2018ljd}, but making use of a WDM cosmology and of more recent data sets.
We finally discuss the results of our SBI analysis in Sec.~\ref{sec:results}. We begin by commenting on the parameter reconstruction capabilities of our procedure, before discussing their degeneracies. In particular, from  Fig.~\ref{fig:triangle-reduced} we can corroborate our expectations that low thresholds for star formation galaxies are a necessary ingredient to probe the WDM scenario in the low mass regime. This is furthermore quantified in our conclusion plots of Figs.~\ref{fig:sensitivitySBI} and \ref{fig:sensitivitySBIb2}, where we display the sensitivity of a fully built radio interferometer, namely HERA, to $\mwdm$ as a function of $\Mt$, assuming a single population of galaxies and 1000h of observations. From these plots, we see that the lower bound prospects on the WDM mass obtained in this work are stronger than the existing bound coming from Lyman-$\alpha$ forest data, provided that the threshold mass $M_{\rm turn }\lesssim 10^8\, M_\odot$. Although this result is fairly robust with respect to the X-ray source properties considered in this study, these properties may influence the strength of the WDM constraint for lower threshold masses. 

As a final comment on the astrophysics modelling, the galaxy model considered here is rather simple and certainly does not account for multiple cosmic dawn galaxy populations that are expected to be present in the early universe. Yet, accounting for the latter will introduce a larger number of parameters that will display extra degeneracies with the WDM mass that should carefully be treated. From our analysis, it is clear that the assumed properties of molecular cooling galaxies, that are expected to be hosted in minihaloes with lower $M_{\rm turn }$, will play a decisive role in the 21cm sensitivity to NCDM properties. Similar conclusions were already drawn in e.g.~\cite{Giri:2022nxq} considering a different analysis and 21cm signal implementation. Our work thus advocates for clearly specifying both the NCDM and astrophysics implementations when reporting prospective constraints on NCDM models with 21cm Cosmology.

\acknowledgments

 We thank G. Facchinetti and S. Palomares-Ruiz for useful discussions. LLH and QD have been supported by the Fonds de la Recherche Scientifique F.R.S.-FNRS  through a research associate position and a FRIA PhD grant, and acknowledge support of the  FNRS research grant number J.0134.24, the ARC program of the Federation Wallonie-Bruxelles and the IISN convention No. 4.4503.15.
Computational resources have been provided by the Consortium des Equipements de Calcul Intensif (CECI), funded by the Fonds de la Recherche Scientifique de Belgique (F.R.S.-FNRS) under Grant No. 2.5020.11 and by the Walloon Region of Belgium. AD and BZ acknowledge the support from CIDEGENT/2020/055. This work was supported by Generalitat Valenciana grant
PROMETEO/2021/083, and by Spanish Ministerio de Ciencia e Innovacion grant PID2020-
113644GB-I00. We also thank the computer resources at IFIC; in particular the SOM cluster, nowadays funded through the grants EUR2022-134028 and ASFAE/2022/020, as well
as Artemisa, funded by the European Union ERDF and Comunitat Valenciana as well as
the technical support provided by the Instituto de Física Corpuscular, IFIC (CSIC-UV),
Generalitat Valenciana, Project “ARTEMISA”, ref. ASFAE/2022/024.

\appendix


\section{Some extra information}
\label{app:moreMock}

\begin{figure}[t!]
\centering
\includegraphics[width=0.99\linewidth]{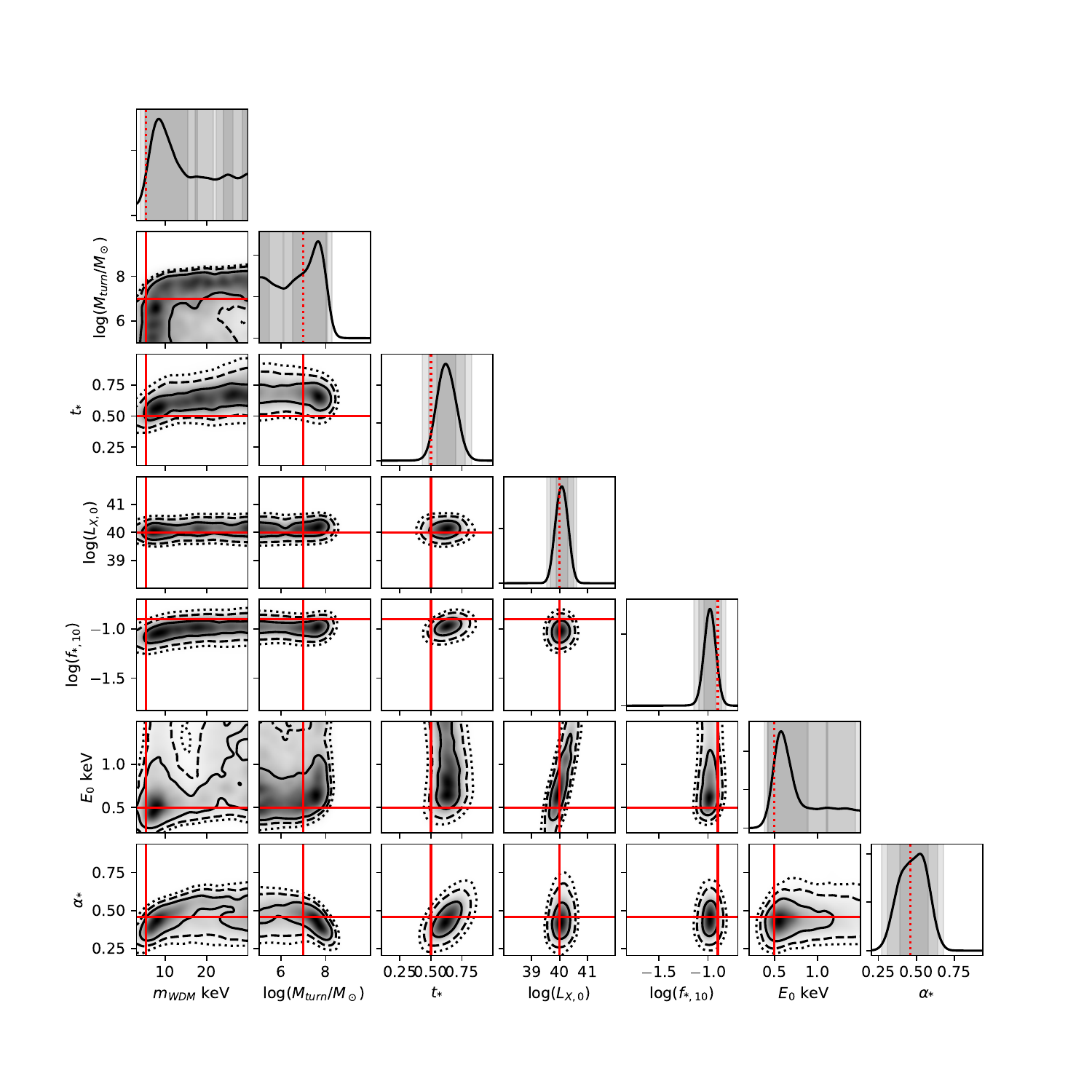}
\caption{Recovered 1D and 2D marginalized posteriors on all astrophysics and WDM  parameters considered here assuming a mock 1000h observation of the 21-cm power spectrum with HERA for $\mwdm=5.3$~keV and $M_{\rm turn}=10^7 M_\odot$. The other parameters have been fixed to the same fiducial values as in Fig.~\ref{fig:PSnonoise}. A reduced version of this plot can be found in Fig.~\ref{fig:triangle-reduced}.}
\label{fig:triangle-full-9}
\end{figure}

\begin{table}[ht!]
\centering
\begin{tabular}{|c||c|c|c||c|c|c|c|}
\hline
Bin & $z_{\rm min}$ & $z_{\rm max}$ & $z_{\rm center}$ & $k_{\rm min}$ & $k_{\rm max}$ & $k_{\rm center}$ \\
\hline
1  & 5.9  & 6.2  & 6.0  & 0.12 & 0.18 & 0.15 \\
2  & 6.2  & 6.5  & 6.3  & 0.18 & 0.24 & 0.21 \\
3  & 6.5  & 6.8  & 6.6  & 0.24 & 0.30 & 0.27 \\
4  & 6.8  & 7.2  & 7.0  & 0.30 & 0.36 & 0.33 \\
5  & 7.2  & 7.6  & 7.4  & 0.36 & 0.42 & 0.39 \\
6  & 7.6  & 8.0  & 7.8  & 0.42 & 0.48 & 0.45 \\
7  & 8.0  & 8.5  & 8.2  & 0.48 & 0.54 & 0.51 \\
8  & 8.5  & 9.0  & 8.7  & 0.54 & 0.60 & 0.57 \\
9  & 9.0  & 9.6  & 9.3  & 0.60 & 0.66 & 0.63 \\
10 & 9.6  & 10.3 & 9.9  & 0.66 & 0.72 & 0.69 \\
11 & 10.3 & 11.0 & 10.6 & 0.72 & 0.78 & 0.75 \\
12 & 11.0 & 11.9 & 11.5 & 0.78 & 0.84 & 0.81 \\
13 & 11.9 & 12.9 & 12.4 & 0.84 & 0.90 & 0.87 \\
14 & 12.9 & 14.1 & 13.5 & 0.90 & 0.96 & 0.93 \\
15 & 14.1 & 15.5 & 14.8 & 0.96 & 1.02 & 0.99 \\
16 & 15.5 & 17.2 & 16.3 & --   & --   & --   \\
17 & 17.2 & 19.3 & 18.2 & --   & --   & --   \\
18 & 19.3 & 21.9 & 20.5 & --   & --   & --   \\
19 & 21.9 & 25.3 & 23.5 & --   & --   & --   \\
\hline
\end{tabular}
\caption{Table of $z$ and $k$ (in units of $[{\rm Mpc}^{-1}]$) bins  edges with their centers.}
\label{tab:zk_bins}
\end{table}

In this appendix we first provide the tabulated values of the considered $z$ and $k$ values in our analysis as described in  Sec.~\ref{sec:Mock}, see Tab.~\ref{tab:zk_bins}. We also provide in Fig.~\ref{fig:triangle-full-9} the full set of recovered 1D and 2D marginalized posteriors on our astrophysics
and WDM parameters. We have assumed a mock 1000h observation of the 21-cm power spectrum with HERA for $\mwdm = 5.3$ keV and $\Mt= 10^7 \rm M_\odot$ and all other parameters set to the same values as in Fig.~\ref{fig:PSnonoise}. As further discussed in Sec.~\ref{sec:2Dposteriors}, for this choice of fiducial, we see that except for $\Mt$, $\as$, $\ts$ and $\fst$, no other astrophysics parameter shows significant degeneracies with $\mwdm$.

\section{Coverage tests }
\label{app:coverage}

\begin{figure}[t!]
\centering
\includegraphics[width=0.49\linewidth]{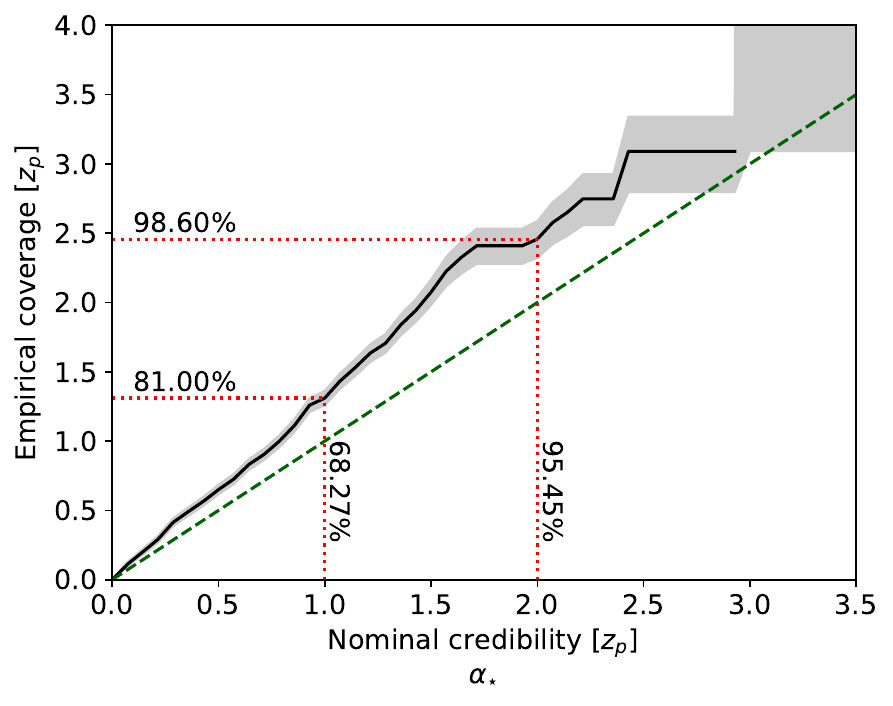}
\includegraphics[width=0.49\linewidth]{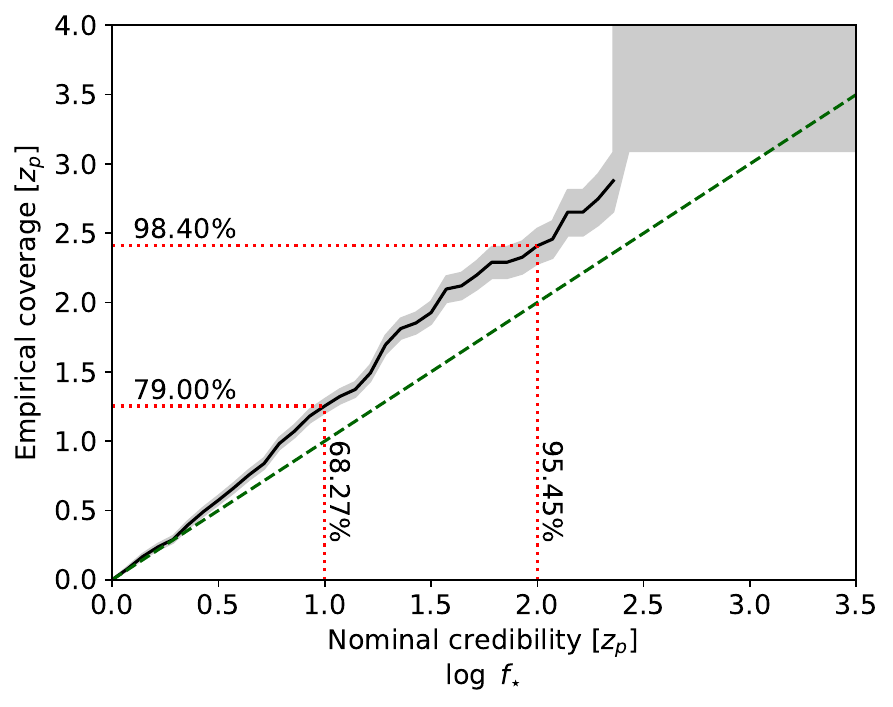}
\includegraphics[width=0.49\linewidth]{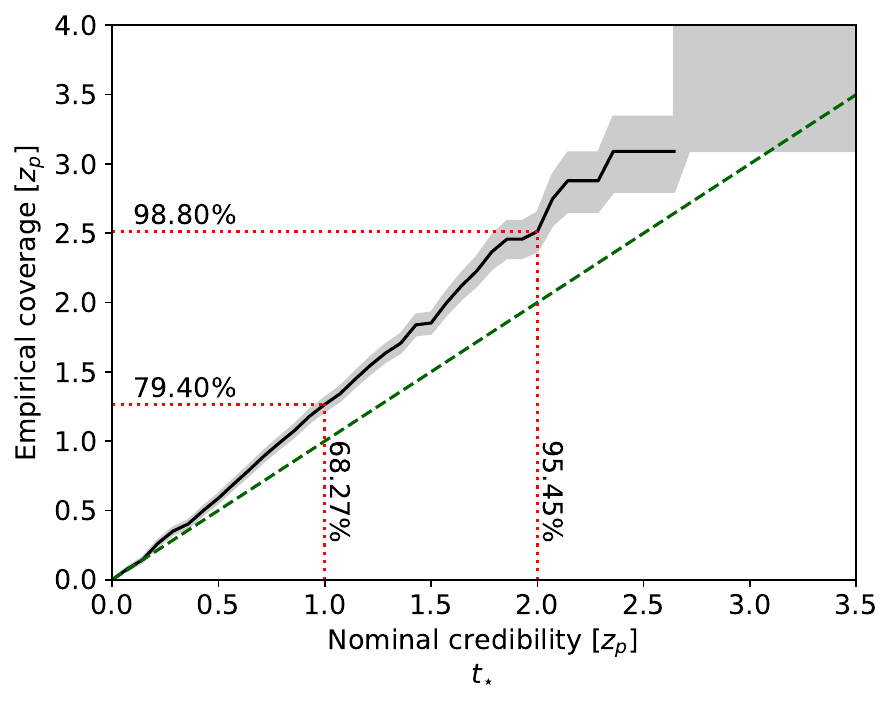}
\includegraphics[width=0.49\linewidth]{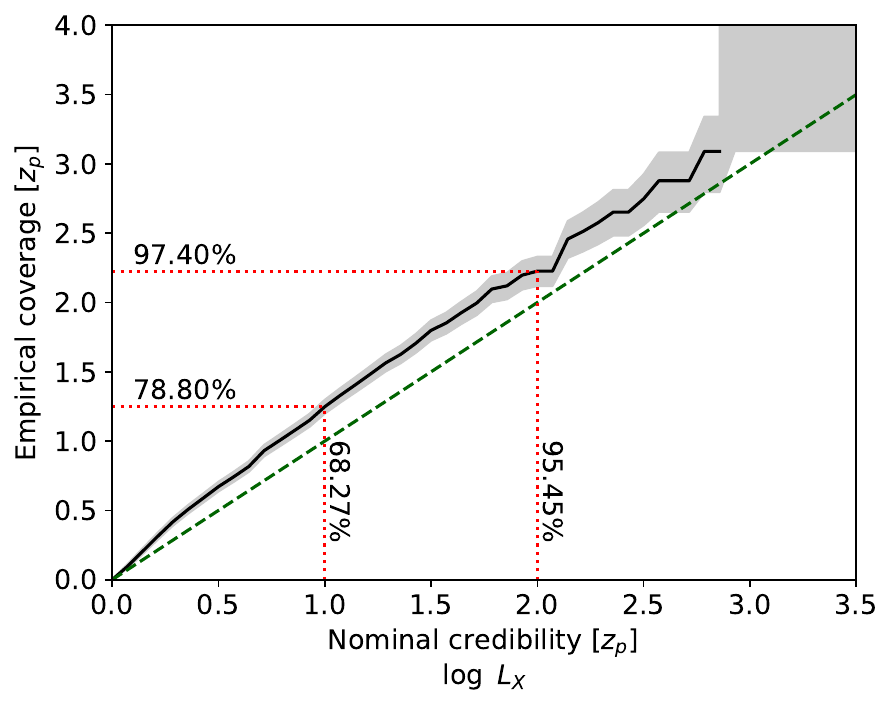}
\includegraphics[width=0.49\linewidth]{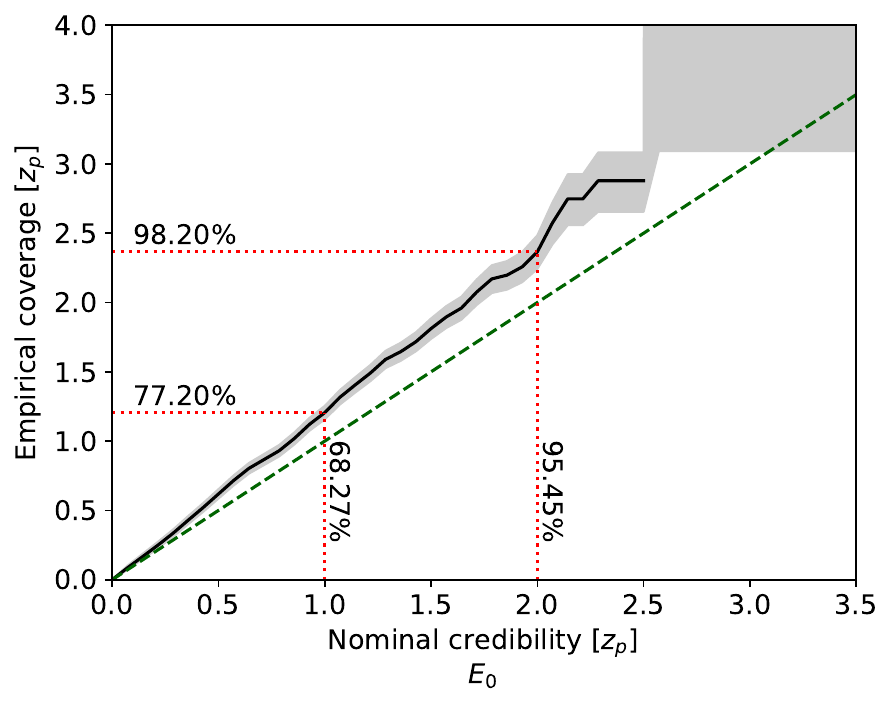}

\caption{Here we follow \cite{2022JCAP...09..004C} displaying the empirical expected coverage probability as a function of the confidence level, $1-\alpha =0.6827,0.9545,0.9997$, or equivalently $z=1,2,3$, for all the astrophysical parameters (except $M_{\rm turn}$). The vertical dashed red lines represent the confidence levels, while the horizontal lines depict the empirical estimates.}
\label{fig:coverage}
\end{figure}

\begin{figure}[t!]
\centering
\includegraphics[width=0.49\linewidth]{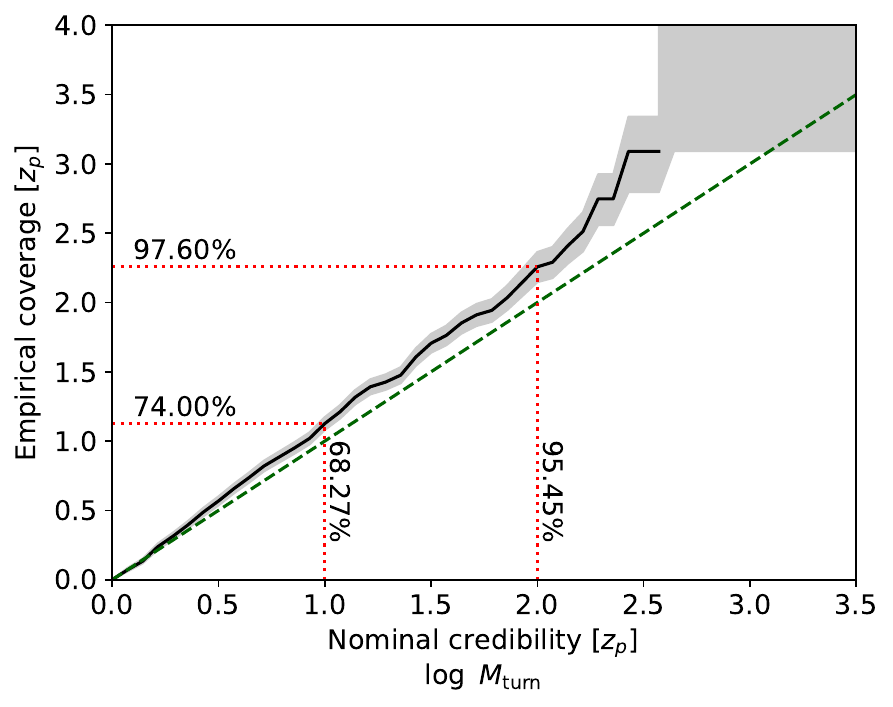}
\includegraphics[width=0.49\linewidth]{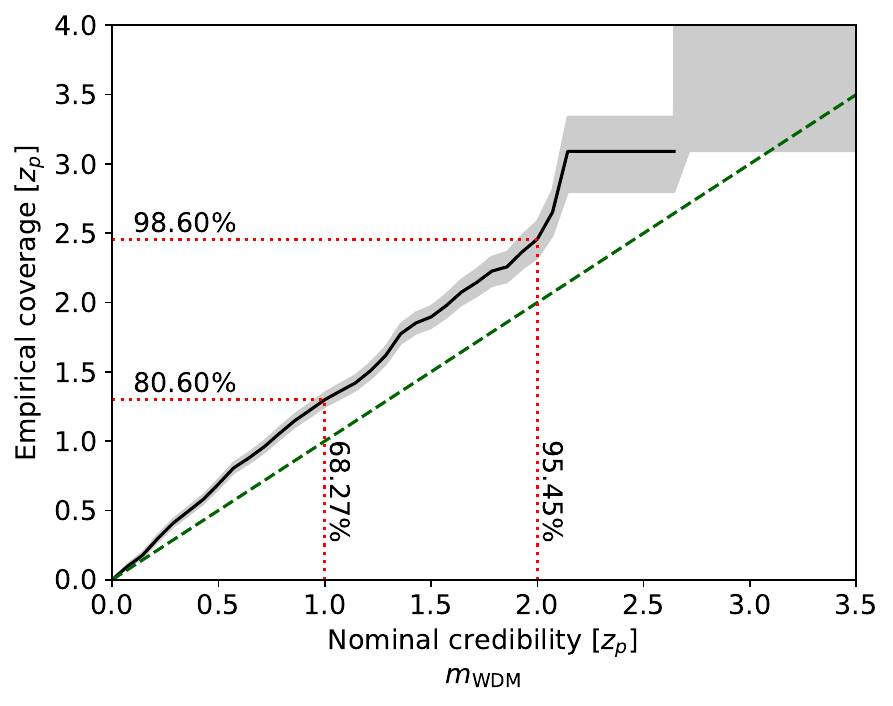}
\caption{Same as Fig. \ref{fig:coverage}, but for $M_{\rm turn}$ and $\mwdm$ parameters.}
\label{fig:coverage2}
\end{figure}
In this appendix, we explore methods to evaluate the quality of the approximate posterior distributions generated by our approach, without relying on MCMC comparison.

Assessing the reliability of posterior inference in simulation-based methods is an active field of research \cite{2021arXiv211006581H}. In this work, we focus on one of the most widely used techniques, commonly referred to as coverage tests or “percentile-percentile” plots. The underlying concept is straightforward: for a given parameter, the x\% credible interval (CI) of the estimated posterior distribution should encompass the true parameter value in approximately x\% of the cases (empirical coverage, EC), when inference is repeated across multiple simulated datasets. Further discussion of this idea can be found in \cite{2021arXiv211006581H}. A well-calibrated posterior would produce an empirical coverage curve that aligns closely with the diagonal line EC=CI in a EC versus CI plot.

Figures~\ref{fig:coverage}  and~\ref{fig:coverage2} present our coverage plots for all parameters considered in our analysis. The gray error bands correspond to the range of the empirical coverage variable, which follows a beta distribution, derived from the probability-point function (PPF). Ideally, the curves in these plots should lie near the diagonal. If the empirical coverage is larger than the nominal credibility (thus, being above the diagonal), the credible intervals are conservative, meaning they capture the true value more often than expected. On the contrary, if the empirical coverage is smaller than the nominal credibility (thus being below the diagonal), the intervals are overconfident and capture the true value less frequently than expected. According to~\cite{2021arXiv211006581H}, a posterior estimator is deemed acceptable when the empirical coverage probability meets or exceeds the specified confidence level.

The results in Figs.~\ref{fig:coverage} and~\ref{fig:coverage2} indicate that the network has converged with reasonable coverage, as the empirical expected coverage probabilities are close to (but above) the corresponding confidence levels, so obtaining relatively conservative posteriors.

\addcontentsline{toc}{section}{Bibliography} 
\bibliographystyle{JHEP}
\bibliography{references} 

\end{document}